%% file: paper.tex
\documentclass[submission, Phys, dvipsnames]{SciPost}
\input{packages}
\input{acronyms}
\input{defs}

\input{tikzdefs}

\begin{document}
	
\begin{center}
	{\Large 
		\textbf{
			Predicting critical temperature in quantum simulators for high-$T_c$ superconductivity: the matrix product state plus mean field approach
		}
	}
\end{center}

\begin{center}
	T.~K\"ohler\textsuperscript{1},
	A.~Kantian\textsuperscript{1}
\end{center}

\begin{center}
	{\bf 1} 
	SUPA, Institute of Photonics and Quantum Sciences, Heriot-Watt University, Edinburgh EH14 4AS, United Kingdom
\end{center}

\section*{\color{scipostdeepblue}{Abstract}}
\textbf{\boldmath{%
	Quantum simulation based on ultra cold atomic lattice gases is one of the most promising platforms to investigate high\hyp $T_c$ superconductivity beyond the limited capabilities of quantum many body numerics on classical computers.
	Yet, despite enormous progress since the field's inception, realizing a high\hyp $T_c$ superconducting state still remains out of reach.
	The present work lays the groundwork to purpose the recently proposed, and already partly realized, mixed-dimensional (mixD) models, towards this end.
	These systems offer the proven capability to realize very high pairing energies while retaining appreciable mobility of pairs.
	We specifically investigate the potential of 2D mixD\hyp models with anisotropic tunneling, using the matrix product state plus mean field theory (MPS+MF) for fermions, and show that these models may enter a high\hyp $T_c$ superconducting phase.
	These simulations in turn are based on a comprehensive characterization of the 1D mixD systems, which are the sub-units of which the 2D system is comprized.
	In this, we cover the range of currently experimentally relevant system sizes, and establish practical heuristics to determine when finite\hyp size effects preclude the use of a 1D mixD\hyp system to build the 2D ones.
}}

\vspace{\baselineskip}

\section{Introduction}%
\Gls{highTc} superconductivity is a major subject of study in the theory of correlated quantum matter, spanning the areas of solid state and condensed matter physics, ultra cold atomic gases and many\hyp body numerics. 
Much early and current theory attempting to explain \gls{highTc} superconducting states in the cuprates and beyond centred on one of several different versions of the 2D Hubbard model~\cite{Schulz1987,Lederer1987,Kotliar1988,Inui1988,Scalapino1995,Furukawa1998,Orenstein2000,Metzner2006,Maier2006b,Raghu2010,Anderson2013,Maier2005,chang_spin_2010,leblanc_solutions_2015,Zheng2017,qin_absence_2020,Xu2022,Xu2024}.
The apparent simplicity of this group of models has proven to be deceptive: these turned out to be highly challenging for arriving at a widely accepted theory of their low\hyp temperature physics.
This applies especially to the problem of predicting which model, if any, realizes a $d$\hyp wave superconducting phase at low or zero temperature, or whether instead one of the multiple other possible correlated phases, which may be insulators or poor conductors, wins the competition between these wildly different ordered states.
The outcome of these competitions is regularly decided by small, or even minute shifts in the parameter values of the microscopic model.
The recent advances towards clarifying the ground\hyp state ordering in some 2D Hubbard models based on large\hyp scale many\hyp body numerics~\cite{leblanc_solutions_2015,Zheng2017,qin_absence_2020,Xu2022,Xu2024} also highlight how long the road to being able to deliberately design \gls{highTc} bulk materials based on weakly coupled 2D planes still remains.
Namely, the microscopic mechanism behind repulsively mediated pairing of electrons, the cornerstone of all \gls{highTc} superconductivity (and unconventional superconductivity more broadly), has not been accessible in this way so far, and remains unresolved.
Likewise, the stability of any correlated phase predicted for a specific version of the 2D Hubbard model remains very challenging to evaluate once a 3D model material is formed from stacking many 2D Hubbard models atop one another, as would be the prediction of the properties of the bulk even if the correlated phase were to survive.

Against this background, the present work merges several active strands of research in order to advance the theory of \gls{highTc} superconductivity well beyond what is currently possible for the 2D Hubbard models. 
The first of these strands is the recently demonstrated capability to design model \gls{highTc} superconductors in bulk using the so called \gls{MPSMF} framework for fermions~\cite{Bollmark2023,Marten2023,Bollmark2025}.
This approach builds on the fact that, almost uniquely, in 1D systems such as two\hyp leg Hubbard ladders the microscopic mechanism for the pairing of electrons from repulsive interactions is known and understood, both from analytical~\cite{giamarchi_quantum_2003,karakonstantakis_enhanced_2011} as well as from numerical techniques~\cite{Trebst2006,karakonstantakis_enhanced_2011,dolfi_pair_2015}.
The \gls{MPSMF} approach then allows to model bulk 3D \gls{highTc} superconductors quantitatively that are built from such 1D systems~\cite{Bollmark2023}.
Crucially, this includes the capability of resolving any competition between multiple orders, such as between superconductivity and insulating phases~\cite{Bollmark2025}.
The second strand are the recently proposed so\hyp called \gls{mixD} systems, bilayer systems in 1D and 2D in which single\hyp fermion tunneling is only present within each layer, while inter\hyp layer coupling consists only of spin exchanges~\cite{Bohrdt2021,Hirthe2023,Lange2026}.
Research into these systems is motivated in part for their potential to manifest high paring energies, which makes them interesting platforms for the study of repulsively mediated superconductivity, possibly with a \gls{highTc} character.
Counter\hyp intuitively, 1D \gls{mixD} systems have been argued to show high mobility of paired fermions while simultaneously realizing high pairing energies~\cite{Bohrdt2021}.
Materials such as La$_3$Ni$_2$O$_7$ are thought to be of the \gls{mixD} type~\cite{kaneko_pair_2024,qu_bilayer_2024}, but 1D \gls{mixD} systems have also already been realized in ultra cold atomic gases~\cite{Hirthe2023}.
The latter systems then form the third strand that the present work takes up and fuses together.
Within the last two decades, ultra cold lattice gases have emerged as some of the foremost quantum simulators for the physics of correlated matter~\cite{bloch_many-body_2008}.
A key area of applications for these experiments has been to improve the understanding of the physics of the 2D Hubbard models at low temperature, some of which already go beyond the limitations of today's classical computational many\hyp body theory~\cite{Parsons2016,Cheuk2016,greif_site-resolved_2016,Mazurenko2017,koepsell_imaging_2019,Bohrdt2019,Bohrdt2021a,Xu2025,kendrick_pseudogap_2025}.
For ultra cold gases, entropy rather than temperature is the figure of merit, as the interacting particles in these experiments realize effective microcanonical ensembles, being highly isolated from the environment by design.
But despite impressive recent progress towards this goal~\cite{Xu2025}, the achievable entropies still appear to be too high to realize any of the conjectured correlated phases of the doped 2D Hubbard model at low temperature~\cite{kendrick_pseudogap_2025}.
Thus, the realization of a \gls{highTc} superconducting state within a quantum simulator, a central aim within the field for over two decades, has proven to be elusive so far.

In the present work, we lay the groundwork to make manifest progress towards this important goal by extensively characterizing the 1D \gls{mixD} systems, one version of which has already been realized in ultra cold lattice gases~\cite{Hirthe2023}.
Our overall strategy is to bypass the 2D Hubbard models completely by establishing \gls{mixD} systems in 1D as building blocks of 2D bilayer \gls{mixD} systems, as shown in~\cref{fig:model:full}, with the potential to host \gls{highTc} superconducting states within the present entropy restrictions in quantum simulators based on ultra cold lattice gases.
Crucially, such bilayer systems can be realized with the latest generation of confocal quantum gas microscopes~\cite{Bakr2009,Sherson2010,Cheuk2015a}.
For one, we focus on characterizing those properties of the 1D \gls{mixD} systems that quantify their conjectured ability to sustain superconducting correlations at high temperatures due to the high pairing energies that these systems can reach.
We do so explicitly not for the thermodynamic limit, but for system sizes and boundary conditions that current experiments can actually realize.
Our analysis yields a significantly more complex picture than previously known: while we do find high pairing energies when fermion mobility itself is high, in line with previous work, we show that this comes at the expense of another performance\hyp critical gap, namely the spin gap, which decreases significantly at the same time.
Likewise, we find that while increasing fermion mobility yields higher superconducting stiffness, this directly depresses the superconducting susceptibility.
Applying the \gls{MPSMF} framework, we then take the version of the 1D \gls{mixD} systems that has already been experimentally realized~\cite{Hirthe2023} and form a 2D bilayer system out of many copies of these.
We show that the resulting system would indeed realize a \gls{highTc} superconducting state.
But again, counter to previous predictions, our results indicate that at fixed inter\hyp ladder coupling the $T_c$ of the resulting bilayer systems is highest when the pairing energy is relatively low.
Finally, given that we model the modest sizes of actually feasible experiments, we delineate where in parameter space the assumptions underlying the field theory description of the 1D \gls{mixD} systems, and thus those of the finite temperature \gls{MPSMF} calculations that we deploy here, break down.

\section{The Model}%
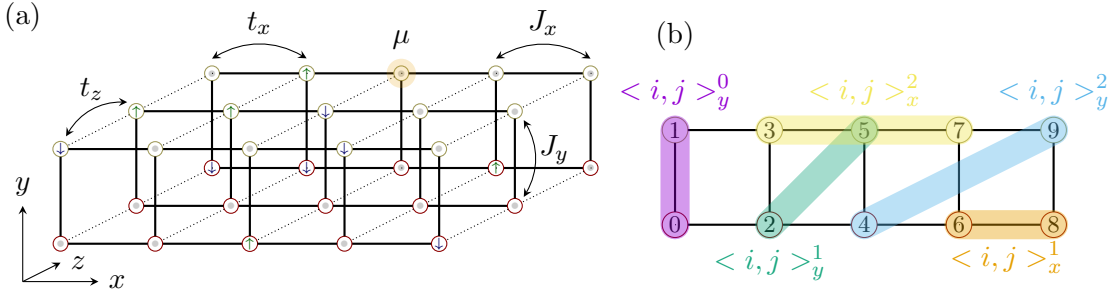
\begin{figure}%
	\begin{subfigure}[b]{0.55\textwidth}
		\centering%
		\ifthenelse{\boolean{buildtikzpics}}%
		{%
			\tikzsetnextfilename{model}%
			\begin{tikzpicture}%
				\node[] at (-2.5,2) (label) {\phantomcaption\label{fig:model:full}(a)};
				\coordinate (origin) at (-2.5,-1.5);%
				\draw[->] (origin) -- ($(origin)+(0,1)$) node[above] {$y$};%
				\draw[->] (origin) -- ($(origin)+(1,0)$) node[right] {$x$};%
				\draw[->] (origin) -- ($(origin)+(0.5,0.25)$) node[right] {$z$};%
				\SystemsInternal{5}{2}{1}{1.0}{\tJs{0.5}{green!40!black}{blue!40!black}{black!20!white}}{1}{3}{1}{1}%
				\draw[<->, bend left] ($(010.north)+(0,0.25em)$) to node[above]  {$t_x$} ($(110.north)+(0,0.25em)$);%
				\node[fit = (210), opacity = 0.2, fill = colorD, circle, inner sep = 0.1em] () {};%
				\node[anchor=south] at ($(210.north)+(0,0.25em)$) {$\mu$};%
				\draw[<->, bend left] ($(310.north)+(0,0.25em)$) to node[above]  {$J_x$} ($(410.north)+(0,0.25em)$);%
				\draw[<->, bend right] ($(011.north west)+(-0.125em,0.125em)$) to node[above]  {$t_z$} ($(012.north east)+(0,0.25em)$);%
				\draw[<->, bend left] ($(411.south east)+(0.125em,-0.125em)$) to node[right, yshift=0.3em, xshift=-0.2em]  {$J_y$} ($(401.north east)+(0.125em,0.125em)$);%
			\end{tikzpicture}%
		}%
		{%
			\phantomcaption\label{fig:model:full}%
			\includegraphics{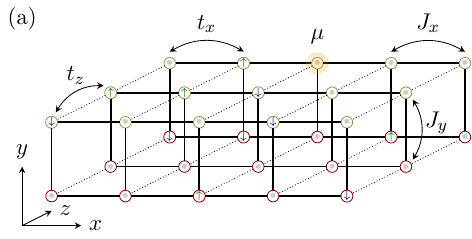}%
		}%
	\end{subfigure}%
	\begin{subfigure}[b]{0.45\textwidth}%
		\centering%
		\ifthenelse{\boolean{buildtikzpics}}%
		{%
			\tikzsetnextfilename{model1D}%
			\begin{tikzpicture}%
				\SystemsInternal{5}{2}{1}{1.0}{{\footnotesize\,\pgfmathparse{int(2*\x+\y)}\pgfmathresult\,}}{1}{1}{1}{1}%
				\node[minimum height=5em, minimum width=4em] {};
				\node[] at (0,2.5) (label) {\phantomcaption\label{fig:model:ladder}(b)};
				\draw[line width=1em, draw opacity=0.4, draw=colorA,line cap=round] (000.center) -- (010.center);
				\node[above = -0.1em of 010, colorA] {$<i,j>^0_y$};
				\draw[line width=1em, draw opacity=0.4, draw=colorB,line cap=round] (100.center) -- (210.center);
				\node[below = -0.1em of 100, colorB] {$<i,j>^1_y$};
				\draw[line width=1em, draw opacity=0.4, draw=colorC,line cap=round] (200.center) -- (410.center);
				\node[above = -0.1em of 410, colorC] {$<i,j>^2_y$};
				\draw[line width=1em, draw opacity=0.4, draw=colorD,line cap=round] (300.center) -- (400.center);
				\node[circle, inner sep = 0pt, outer sep = 0pt] (tmp) at ($(300)!0.5!(400)$) {$\vphantom{1}$};
				\node[below = -0.1em of tmp, colorD] {$<i,j>^1_{x\vphantom{y}}$};
				\draw[line width=1em, draw opacity=0.4, draw=colorE,line cap=round] (110.center) -- (310.center);
				\node[above = -0.1em of 210, colorE] {$<i,j>^2_{x\vphantom{y}}$};
			\end{tikzpicture}%
		}%
		{%
			\phantomcaption\label{fig:model:ladder}%
			\includegraphics{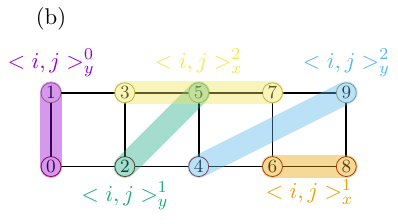}%
		}%
	\end{subfigure}
	\caption%
	{%
		\label{fig:model}
		\subfigref{fig:model:full} 
		Overview of the studied tJ\hyp model describing the 2D \gls{mixD} system, viewed as an array of coupled ladders stacked in the $z$\hyp direction, with tunneling between them given by $t_z$.
		Each ladder extends in the $x$\hyp $y$\hyp plane, and has intra\hyp layer tunneling $t_x$ and an effective inter\hyp layer tunneling $t_y=0$, while inter\hyp layer spin exchange $J_y$ is the largest scale and used as the unit of energy (cf. main text).
		\subfigref{fig:model:ladder} Effective model solved within the \gls{MPSMF} framework, the t\hyp J Hamiltonian for a single 1D \gls{mixD} system exposed to several self\hyp consistent mean\hyp field amplitudes mimicking the full 2D array in the regime ${t_z<E_p,2E_s}$.
		Indices indicate the order in which the tensors inside the \gls{MPS} are arranged.
	}%
\end{figure}%
The principle underlying the realization of \gls{mixD} systems in ultra cold atomic lattice gases is that of metastable many\hyp body state preparation for fermions~\cite{Rabl2003,kantian_atomic_2007,Kantian2010}, where an effective pseudospin\hyp $1/2$ is realized via two equally populated atomic hyperfine states:
one of the layers of the bilayer optical lattice that is trapping the atoms is realized with a substantial energy offset relative to the other layer, as large as half of the repulsive on\hyp site interaction $U$, the largest scale within the underlying microscopic single-band Hubbard model.
When prepared in this way at equal density with a large barrier between the layers, exceptionally strong inter\hyp layer spin exchange coupling can be realized. 
This is achieved by slowly decreasing the barrier, while maintaining the energy offset between the layers. 
This offset further suppresses any effective inter\hyp layer single\hyp fermion tunneling~\cite{Hirthe2023}
From the Hubbard model, one can therefore derive an effective t\hyp J type Hamiltonian for the physics of the metastable state\hyp manifold of this system, and which has further been demonstrated experimentally to be well\hyp suited.
We start the present work from such an effective Hamiltonian for the two coupled planes, shown in~\cref{fig:model:full}.
With an eye already towards approximation within the \gls{MPSMF}\hyp framework, we write the Hamiltonian in the following form:
\begin{align}\label{eq:fullH}
	\hat H = \hat H_{\mathrm{tJ-ladder}} + \hat H_{\mathrm{inter ladder}}\;.
\end{align}
Here, $\hat H_{\mathrm{tJ-ladder}}$ describes the t\hyp J ladders, stacked in the $z$-direction and each of length $L_x$ in the $x$\hyp direction, making up the 2D bilayer.
The term $\hat H_{\mathrm{inter ladder}}$ then describes the tunneling in $z$\hyp direction, i.e., along the stack of ladders.
Thus, the first term on the right side of~\cref{eq:fullH} contains the chemical potential $\mu$, the \gls{NN} $x$\hyp direction tunneling $t_x$, while $y$\hyp direction tunneling $t_y=0$, as well as the spin\hyp exchange coupling $J_d$ where the direction is indicated by $d=x,y$ (cf. \cref{fig:model}):
\begin{align}\label{eq:tJladder}
	\hat H_{\mathrm{tJ-ladder}} =& \sum_j -\mu \hat n_j + \sum_{d \in \{x, y\}} \left[ -t_d \sum_{\sigma, <i, j>_d} \left( \hat c^{\dagger}_{i, \sigma} \hat c^\nodagger_{j, \sigma} + \mathrm{h.c.} \right) \right. \nonumber\\
	&\left.+ J_d \sum_{<i,j>_d} \left(\frac{1}{2}(\hat S^+_i \hat S^-_{j} + \hat S^-_i \hat S^+_{j}) + \hat S^z_{i} S^z_{j} + \frac{1}{4} \hat n_i \hat n_{j}\right) \right]\;,
\end{align}
Here, indices $i$ and $j$ run over all sites of the 2D bilayer system.
In these models, as realized in ultracold atomic lattice gases, $J_x\ll J_y$~\cite{Hirthe2023}.
In fact, $J_y$ can be so large as to represent the largest coupling in the problem, which is why we use it as the unit of energy in the following, i.e., $J_y=1$.
As usual, summation over \gls{NN} pairs in direction $d$ is denoted by $<i,j>_d$, and the two populated hyperfine states of the atoms are encoded as the pseudospin\hyp $1/2$, $\sigma=\uparrow,\downarrow$.
With this notation, the second term on the right side of \cref{eq:fullH}, the tunneling between the ladders along the $z$\hyp direction of the stack, is expressed as
\begin{align}
	\hat H_{\mathrm{inter ladder}} =
		& - t_z \sum_{\sigma, <i,j>_z} 
		\left( 
			\hat c^{\dagger}_{i,\sigma} \hat c^{\nodagger}_{j,\sigma} 
			+ \mathrm{h.c.} 
		\right)
\end{align}
The \gls{MPSMF}\hyp framework then allows us to derive an effective single\hyp ladder Hamiltonian from the full Hamiltonian \cref{eq:fullH} (cf.~\cref{fig:model:ladder}):
\begin{align}\label{eq:heff}
	\hat H_{\mathrm{effective}} =& \hat H_{\mathrm{tJ-ladder}} - \hat H_{\mathrm{MF}}\;,
\end{align}
Here, $\hat H_{\mathrm{tJ\hyp ladder}}$ is again given by \cref{eq:tJladder}, with the one change that the indices $i$ and $j$ are now running over the sites of a single t\hyp J ladder, as opposed to all the sites of the 2D bilayer system, as they did in \cref{eq:tJladder}.
The action of all the other t\hyp J ladders in the stack on any given one ladder are encoded in the mean\hyp field Hamiltonian:
\begin{align}\label{eq:hmf}
	\hat H_\mathrm{MF} = 
		& 	\sum_{R=0}^{R_{\mathrm{max}}} 
				\sum_{d \in \{x, y\}}
					\left[
						\alpha^R_d \sum_{\sigma, <i,j>^R_d} \left(\hat c^\nodagger_{i,\sigma} \hat c^\nodagger_{j,\sigma} + \mathrm{h.c.}\right)
						+
						\beta^R_d \sum_{\sigma, <i,j>^R_d} \left(\hat c^\dagger_{i,\sigma} \hat c^\nodagger_{j,\sigma} + \mathrm{h.c.}\right)
					\right]
			\;,
\end{align}
In this Hamiltonian, the self\hyp consistently determined mean field amplitudes are given by two families of terms, the \gls{pp} amplitudes $\alpha^R_d$ and the \gls{ph} amplitudes $\beta^R_d$:
\begin{align}\label{eq:amplitudes}
	\alpha^R_d & = 
					\frac{2t_z^2}{E_p} 
					\left\langle
						\sum_{\sigma, <i,j>^R_d}
							\left(
								\braket{\hat c^\nodagger_{i, \sigma}\hat c^\nodagger_{j,\sigma}}
								+\braket{\hat c^\dagger_{i, \sigma}\hat c^\dagger_{j,\sigma}}
							\right)
					\right\rangle \\
	\beta^R_d & = 
					\frac{2t_z^2}{E_p} 
					\left\langle
						\sum_{\sigma, <i,j>^R_d}
							\left(
								\braket{\hat c^\dagger_{i, \sigma}\hat c^\nodagger_{j,\sigma}}
								+\braket{\hat c^\nodagger_{i, \sigma}\hat c^\dagger_{j,\sigma}}
							\right)
					\right\rangle\;.
\end{align}
Here, the outer brackets indicate an average over absolute position, and $<i,j>^R_d$ denotes the set of pairs of sites with distance $R$ between them in the $x$\hyp direction, either on the same leg of the ladder ($d=x$), or on two different legs ($d=y$), as illustrated in \cref{fig:model:ladder}.
We use the position\hyp average here in order to avoid misalignment between the numerical data, which might show some boundary effects otherwise, and the analytical theory used to compute $T_c$ (cf. \cref{sec:method}), which was developed for isotropic systems in the thermodynamic limit.
As all the finite-size data shown in the following indicates, due to the strong pairing energies in these systems these boundary effects are marginal and disappear quickly, and the use of the averages is as good as any other approach for these systems.
Also, we note that $\beta^0_x$ is fixed to zero, because it would otherwise interfere with the chemical potential and it would be impossible to fix a density.

As detailed in refs.~\cite{Bollmark2023,Bollmark2025}, the validity of moving from the original model \cref{eq:fullH} to the approximate one, \cref{eq:heff} with \cref{eq:hmf} and \cref{eq:amplitudes}, rests on the inter\hyp ladder coupling $t_z$ being smaller than either the so\hyp called pairing gap $E_p$ and -- in the case of ultra cold atoms -- twice the so\hyp called spin gap $E_s$ of an isolated ladder.
Both of these are defined in the following section, along with some other quantities that both quantify the ability of the isolated ladder to host superconducting correlations, and that are at the same time required for the efficient calculation of the critical temperature $T_c$ of the 2D bilayer array within the \gls{MPSMF} approach.

\section{Method} \label{sec:method}
\begin{figure}[!th]%
	\centering%
	\ifthenelse{\boolean{buildtikzpics}}%
	{%
		\include{figures/PaP}%
	}%
	{%
		\includegraphics{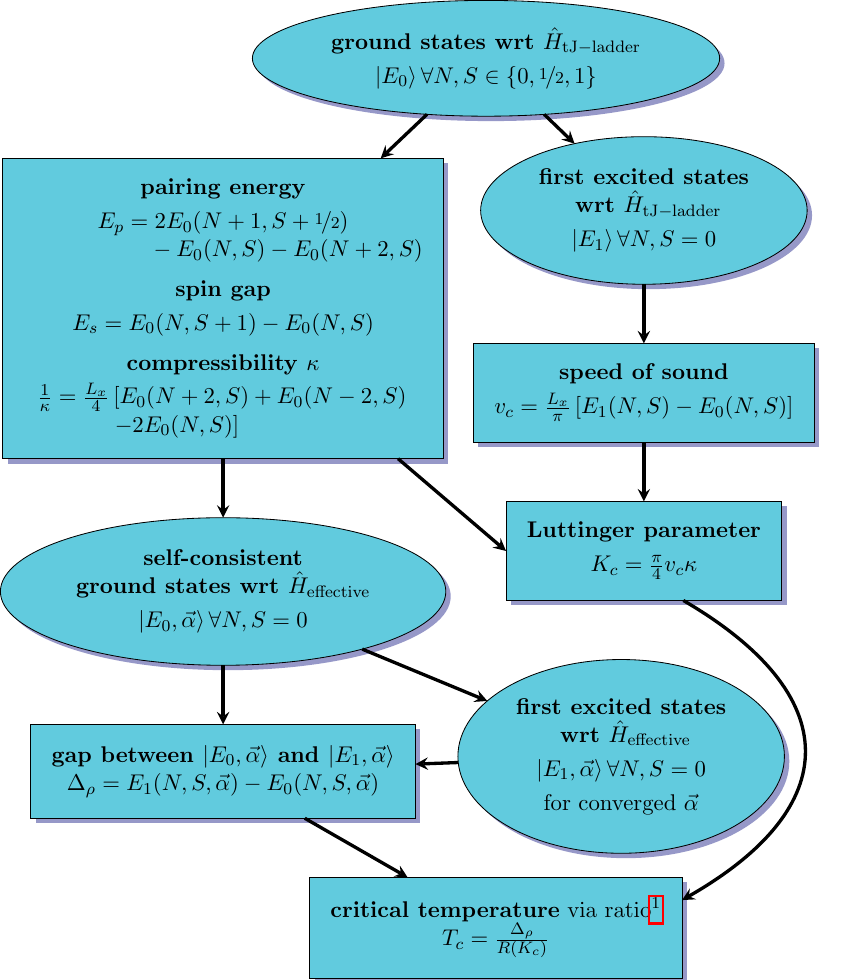}%
	}%
	\caption%
	{%
		\label{fig:pap}%
		Necessary steps to obtain critical temperatures as described in \cref{sec:method}. %
	}%
\end{figure}%
The workflow to obtain the critical temperature $T_c$ for the model in \cref{eq:heff}, through a mix of numerics and analytics within the \gls{MPSMF} framework, is summarized in \cref{fig:pap}. 
This semi\hyp analytical way rests on computing the \gls{TLL} parameter $K_c$~\cite{giamarchi_quantum_2003} of an isolated ladder at the targeted density, as well as calculating both the ground state and the first excited state of the Hamiltonian in \cref{eq:heff} with converged mean-field amplitudes, \cref{eq:amplitudes}, at matching density.
Typically, one knows or expects the superconductor to be in a specific spin-sector $S$, which \gls{MPS}\hyp based numerics can deliberately target in general~\cite{schollwock_schollwock_2011}.
In the present case, there is every reason to expect any superconducting solution to be of singlet\hyp type, and we thus target the sector $S=0$.

\begin{figure}%
	\centering
	\ifthenelse{\boolean{buildtikzpics}}%
	{%
		\input{figures/EpEsCompSos}
	}%
	{%
		\includegraphics{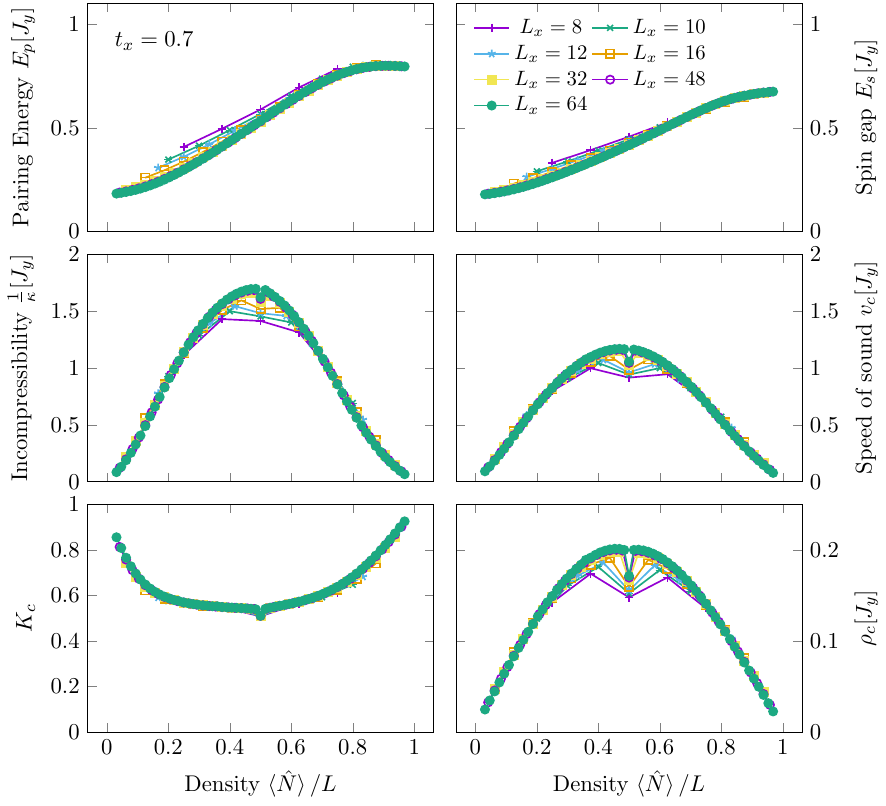}%
	}%
	\caption%
	{%
		\label{fig:EpEsCompSos}%
		Finite size scaling of the six quantities, pairing energy ($E_p$), spin gap ($E_s$), compressibility $(\frac{1}{\kappa}$), speed of sound ($v_c$), Luttinger parameter $K_c$, and superconducting stiffness $\rho_c$ obtained for the isolated t\hyp J ladder at $t_x=0.7J_y$.
		For systems larger than $L_x=32$ the differences are smaller than the symbols used and can be neglected.
	}%
\end{figure}

The first practical step is to calculate the energies of the ground and the first excited states in all relevant particle\hyp number sectors $N=2,4,\dots,2L_x$ at ladder\hyp length $L_x$, $E_0(S, N,L_x)$ and $E_1(S, N,L_x)$ respectively, for the isolated ladder, \cref{eq:tJladder}.
Additionally, the energy of ground states in adjacent spin sectors $E_0(S^\prime,N,L_x)$, i.e., $S^\prime = S + \nicefrac12$ and $S^\prime = S + 1$, needs to be obtained.
It is for isolated ladders where \gls{MPS}\hyp based codes allow us to work at both fixed total number of atoms $N$ and fixed pseudospin $S$.
It is only for the second stage, the \gls{MPSMF} calculations, where $N$ will no longer be a conserved quantum number, and the density of the system needs to be fixed on average through the use of the chemical potential term $\mu$ in \cref{eq:tJladder}, as summarized in \cref{fig:pap}.
From all these energies we obtain the pairing energy, \cite{karakonstantakis_enhanced_2011}
\begin{align}%
	E_p(L_x) = 2 E_0(N+1,S+0.5,L_x) - E_0(N,S,L_x) - E_0(N+2,S,L_x)  \;,%
\end{align}%
and the spin gap, \cite{karakonstantakis_enhanced_2011}
\begin{align}%
	E_s(L_x) = E_0(N,S+1,L_x) - E_0(N,S,L_x) \;,%
\end{align}%
at the desired density $n=N/2L_x$.
When modeling the 2D bilayer system with \gls{MPSMF} numerics, this is replaced by $n=\langle\hat{N}\rangle/2L_x$.
These two gaps, which obey $E_s\leq E_p$, quantify the degree to which itinerant pairs of opposite-spin fermions are stable:
$E_p$ is the minimal amount of energy required to break a pair by moving one of the two constituent atoms from one ladder to another ladder.
Likewise, $E_s$ is the minimal amount of energy needed to break a pair by flipping the pseudospin of one of the constituent atoms.
Different from electrons in solid state materials that have actual spin, ultra cold atoms will not spontaneously flip from one hyperfine state to the other on experimental time scales, staying overall in whatever $S$\hyp sector they were prepared in.
Thus, the minimal process required to break pairs in this way is for two atoms to be exchanged between two ladders, which corresponds to a manifold of states that is at least $2E_s$ above the low\hyp energy manifold where all the atoms are paired-up.
The array of energies $E_0(S, N)$ further allows to compute the compressibility $\kappa(L_x)$ at length $L_x$,~\cite{fehske_metallicity_2008}
\begin{align}%
	\frac{1}{\kappa(L_x) }= L_x \frac{E_0(0, 2N+2) + E_0(0, 2N-2) - 2E_0(0,2N)}{4}\;,%
\end{align}%
and the speed of sound $v_c$ at length $L_x$, \cite{fehske_metallicity_2008}
\begin{align}\label{eq:vc}%
	v_c(L_x) = \frac{L_x}{\pi} \left(E_1(0, 2N,L_x) - E_0(0,2N,L_x)\right)\;.%
\end{align}%
The definition of $v_c(L_x)$ makes it especially susceptible to finite\hyp size effects -- as discussed in \cref{sec:heuristic}, the larger $t_x$ is, the larger $L_x$ has to be to avoid accidentally picking the wrong excited state $E_1$ and yielding an unphysical result.
At the same time, we generally find $E_p$, $E_s$, $\kappa(L_x)$ and $v_c(L_x)$, as well as the properties derived from them, to become remarkably stable beyond surprisingly low values of $L_x$ for a wide range of microscopic parameters, an example of which is shown in \cref{fig:EpEsCompSos}.

From $\kappa(L_x)$ and $v_c(L_x)$, one can directly calculate the superconducting stiffness of the isolated ladder,
\begin{align}
	\rho_c(L_x) = \frac{v^2_c(L_x) \kappa(L_x)}{4}  \;.
\end{align}
as well as the \gls{TLL}\hyp parameter $K_c$ at length $L_x$,
\begin{align}
	K_c(L_x) = \frac{\pi}{4} v_c(L_x) \kappa(L_x) \;.
\end{align}
This \gls{TLL}\hyp parameter is the single variable of the ratio function \cite{Bollmark2023},
\begin{align}\label{eq:ratio}
	R(K_c)=2 \pi \left [ \frac{K_c \tan{\left ( \frac{\pi}{2} \frac{1}{8K_c-1}\right )} }{ \kappa^2(K_c)(8K_c - 1) \sin \left ( \frac{\pi}{4K_c} \right )  B^2 \left (\frac{1}{8K_c}, 1 - \frac{1}{4K_c} \right ) }  \right ]^{\frac{2K_c}{4K_c-1}} \sin \left ( \frac{\pi}{8K_c-1}\right ) \;,
\end{align}
in which $B(a,b)$ denotes the beta function, and $\kappa(K_c)$ is a combination of gamma functions given in \cref{app:analytics}.
It is this ratio\hyp function that allows us to compute $T_c$ from the energy gap $\Delta$ between the ground and first excited state of the effective Hamiltonian, \cref{eq:heff}:
\begin{align}\label{eq:delta}
	\Delta = E_1(N,S,L_x,\vec \alpha) - E_0(N,S,L_x,\vec \alpha)\;.
\end{align}
Here, $E_i(N,S,L_x,\vec \alpha)$ denotes the energy of the $i$th excited state of $\hat H_{\mathrm{effective}}$, which is naturally a function of the vector of mean\hyp field amplitudes $\vec \alpha:=(\alpha^R_d,\beta^R_d)_{R=0,\dots,R_{\mathrm{max}},d=x,y}$, which has been converged up to a pre\hyp set tolerance.
The array's critical temperature $T_c$ for the onset of superconductivity can now be computed from the ratio\hyp function $R(K_c)$ and the gap $\Delta$, by the simple formula
\begin{align}\label{eq:tc}
	T_c(L_x)= \frac{\Delta}{R(K_c)}\;.
\end{align}
For complex ladder systems of the type studied here, this is computationally far cheaper than the alternative, as it requires iterating the mean\hyp field amplitudes to self consistency via repeated zero\hyp temperature \gls{DMRG} calculation to obtain $E_0(N,S,L_x,\vec \alpha)$, after which getting $E_1(N,S,L_x,\vec \alpha)$ is straightforwardly done via one additional ground state search with orthogonality constraint.
The alternative to this would be to compute $T_c$ by brute force, by iterating the mean\hyp field amplitudes to self consistency based on direct calculation of the thermal density matrix at various temperatures, and thereby locating when the \gls{pp} amplitudes $\alpha_d^R$ start to become non\hyp zero.
As we have previously shown that both approaches yield very close results~\cite{Bollmark2023}, and have explained the underlying reasons for this through the conformal field theory that these arrays of coupled 1D systems inherit from their constituent elements, we proceed with the more efficient method in the following.
\section{Results}%
\begin{figure}%
	\centering%
	\ifthenelse{\boolean{buildtikzpics}}%
	{%
		\input{figures/pairingenergyANDspingapSurfacePlot}%
	}%
	{%
		\includegraphics{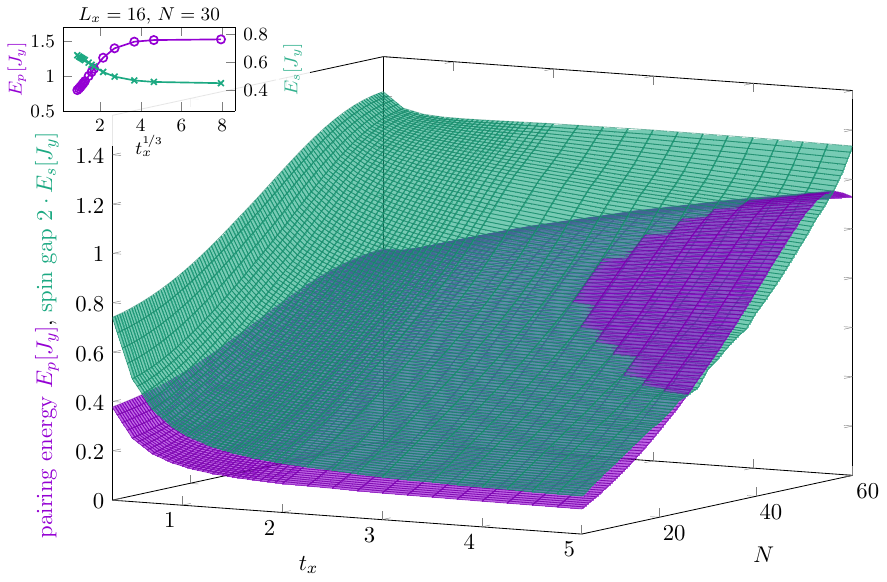}%
	}%
	\caption%
	{%
		\label{fig:EpEsVsTx}%
		Pairing energy and doubled spin gap for $t_x\in [0.3J_y,5J_y ]$ and densities $\nicefrac{N}{2L_x}\in [0,1]$ for an isolated ladder of length $L_x=32$. %
		For high densities, i.e., small hole doping an increase in the pairing energy is shown. %
		Simultanously, the decrease in the spin gap restricts the useful range of $t_x$. %
	}%
\end{figure}%
\begin{figure}%
	\centering
	\ifthenelse{\boolean{buildtikzpics}}%
	{%
			\input{figures/Incompressibility_SoS_K_rho_Lx_32}
	}%
	{%
			\includegraphics{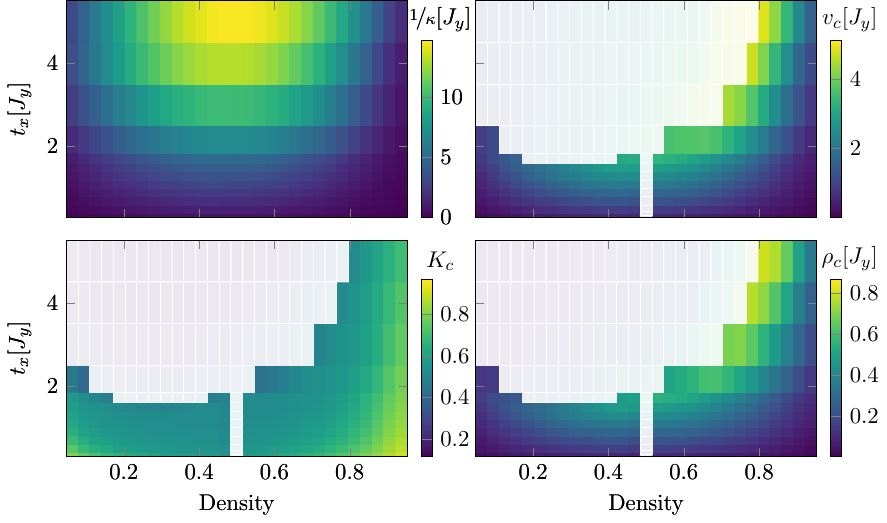}%
	}%
	\caption%
	{%
		\label{fig:Incompressibility_SoS_K_rho_Lx_32}%
		The incompressibility $\nicefrac{1}{\kappa}$, the speed of sound $v_c$, the Luttinger parameter $K_c$, and the helicity modulus $\rho_c$, analog to the four lower subfigures of \cref{fig:EpEsCompSos}.
		Instead of the system size dependency, here the dependency on $t_x$ is shown for $L_x=32$.
		Additionally, for all quantities that depend on the speed of sound, the area in which the speed of sound is ill\hyp defined (cf. \cref{sec:heuristic}) is overlayed by a white fog.
	}%
\end{figure}%
As discussed in the previous section, \cref{fig:EpEsCompSos} illustrates that the key observables quantifying the strength and stability of superconducting correlations in the isolated t\hyp J ladders are affected remarkably little by finite\hyp size effects.
In the following we will therefore mostly focus on the upper range of $L_x$\hyp values for which we present data in \cref{fig:EpEsCompSos}, $L_x=32$, which accord to the linear sizes that can be realized in present quantum gas microscopes.
Here, $E_p$ and $2E_s$ provide stability against one of the two main mechanisms for the loss of superconducting correlations, namely against pair\hyp breaking from either the effects of temperature, or from the effects of additional kinetic energy in the $z$\hyp direction when forming a 2D bilayer system out of many t\hyp J ladders.
Consequently, these gaps set the most important ceilings to the $t_z$\hyp values permitted in the \gls{MPSMF} treatment of the 2D bilayer, as discussed in more detail in \cref{subsec:tc}.
The compressibility $\kappa$ and speed of sound $v_c$ are not just important to calculate the superconducting phase stiffness $\rho_c$ as well as the \gls{TLL} parameter $K_c$, which parametrizes the superconducting susceptibility of the t\hyp J ladder, which diverges as $T^{2-\frac{1}{2K_c}}$~\cite{giamarchi_quantum_2003} and is required for computing $T_c$ of the array, \cref{eq:tc}, via the ratio\hyp function, \cref{eq:ratio}.
The speed of sound is also the measure for the velocity with which excitations in the pair\hyp density spread in the ladder, and thus are one way to quantify the mobility of the pairs.
This is why it generally behaves analogously to $\rho_c$ -- the degree to which the ladder can resist fluctuations in the phases of the paired atoms -- with density and $t_x$.
The loss of pair\hyp phase coherence via phase fluctuations is the other main mechanism by which superconducting correlations in the ladder are depressed.

\subsection{Sweet spot for trade\hyp off between $E_p$ and $E_s$ -- performance metrics of the t\hyp J ladder}\label{sec:sweet}
Before deploying the \gls{MPSMF} framework in order to quantify the $T_c$ for the onset of superconductivity in the 2D bilayer system in the following subsection, it bears discussing the performance of the constituent units of the 2D system, the isolated t\hyp J ladders, with respect to superconducting correlations.
We have computed $E_p$, $E_s$, $\kappa$ and $v_c$ -- and thus also $K_c$ and $\rho_c$ -- for a grid of densities between $0$ and $1$ and $t_x$\hyp values between $0.3J_y$ and $5J_y$, for both $L_x=16$ and $L_x=32$.

A key result is that while we find that the pairing gap $E_p$ -- counter\hyp intuitively -- increases with $t_x$, in line with previous results, but only at densities $n>0.5$.
However, this comes at the expense of the spin gap $E_s$.
As summarized in \cref{fig:EpEsVsTx}, the sweep spot of $t_x$ and density, as far as the central importance of the two gaps in the \gls{MPSMF} framework is concerned, occurs around ${t_x\sim 3.5 J_y}$ and for densities roughly between $0.5$ and $0.75$, where $E_p$ and $2 E_s$ coincide.
The inset in \cref{fig:EpEsVsTx} further illustrates that, as previously predicted, $E_p$ does indeed scale up with $t_x^{1/3}$ before ultimately saturating due to finite\hyp size effects, but that $E_s$ also scales down in the same manner, before saturating itself.
At the same time, we stress that this only applies to settings where the spin excitations are gapped away by $2E_s$ from the low\hyp energy manifold of states, such as in the ultra cold gas setting that this work aims at.
If these excitations are only gapped away by $E_s$, as would be the case, e.g., in solid state systems, due to $E_s\leq E_p$ it is the spin gap alone that will firmly constrain the allowed $t_z$\hyp values as well as the $T_c$\hyp values that can meaningfully be predicted by the \gls{MPSMF} framework.

As remarkable as the string\hyp based chargon pairing~\cite{Bohrdt2021} of these t\hyp J ladders at $n>0.5$ is, with the attendant scaling $E_p\propto t_x^{1/3}$, this survey of both gaps already implies that tuning a 2D bilayer system to exhibit \gls{highTc} superconductivity will be more complex than increasing $t_x$ above the value of the initial experiment~\cite{Hirthe2023}.
This reading of our results is only reinforced when comparing the behavior of $K_c$ and $\rho_c$ in \cref{fig:Incompressibility_SoS_K_rho_Lx_32} with that of $E_p$ as shown in \cref{fig:EpEsVsTx}.
For the finite systems that are the target of the present work, any definition used for computing these will always break down at high values of $t_x$, due to finite\hyp size effects.
The speed of sound $v_c$ is particularly affected by this, as discussed in detail in \cref{sec:heuristic}, leading to parts of the $t_x$\hyp $n$\hyp parameter plane being effectively inaccessible for $v_c$ as well as the derived properties $K_c$ and $\rho_cx$, as indicated in \cref{fig:Incompressibility_SoS_K_rho_Lx_32}.

Distinct patterns emerge in the accessible parts of the parameter space.
Specifically, the maxima of $K_c$ in the $t_x$\hyp $n$\hyp plane, which correspond to the fastest divergence of the superconducting susceptibility of the t\hyp J ladder, coincide with low $t_x$\hyp values and with both with the lowest and highest ranges of the values of $E_p$.
Meanwhile, the superconducting stiffness $\rho_c$ shows behavior that is just opposite to that of $K_c$.
And while the values of $\rho_c$ are peaking around $n=0.5$ and increase with $t_x$, these maxima do not coincide with those of $E_p$.
We find the same general behavior holds for the speed of sound $v_c$.
Thus, while high ratios of $t_x/J_y$ generally increase $E_p$ (at least for $n>0.5$), $\rho_c$ and $v_c$, a high value of $E_p/J_y$ does not maximize either one of these alternative measures for the mobility of the pairs.
And $K_c$, which determines superconducting susceptibility, is completely independent from $E_p/J_y$
Therefore, even for the isolated t\hyp J ladder, a blanket increase of $t_x/J_y$ irrespective of the density will only result in limited improvement on some properties related to strong superconducting correlations ($E_p$, $\rho_c$ and $v_c$), while significantly worsening others ($E_s$ and $K_c$).
Against this complex background we deploy the \gls{MPSMF} framework in the next subsection, in order to quantify the impact that changing the density, and thereby changing $E_p$, $E_s$, $v_c$, $\rho_c$, and $K_c$ has on the $T_c$ for the onset of repulsively\hyp mediated superconductivity the 2D bilayer array.

\subsection{Critical temperature}\label{subsec:tc}
\def\tx{0p7}	
\begin{figure}[!th]%
	\centering
	\def\Lx{16}%
	\def\tperpStr{0p4}%
	\def\tperp{0.4}%
	\centering%
	\ifthenelse{\boolean{buildtikzpics}}%
	{%
		\input{figures/all_Tcs_with_Ep}%
	}%
	{%
		\includegraphics{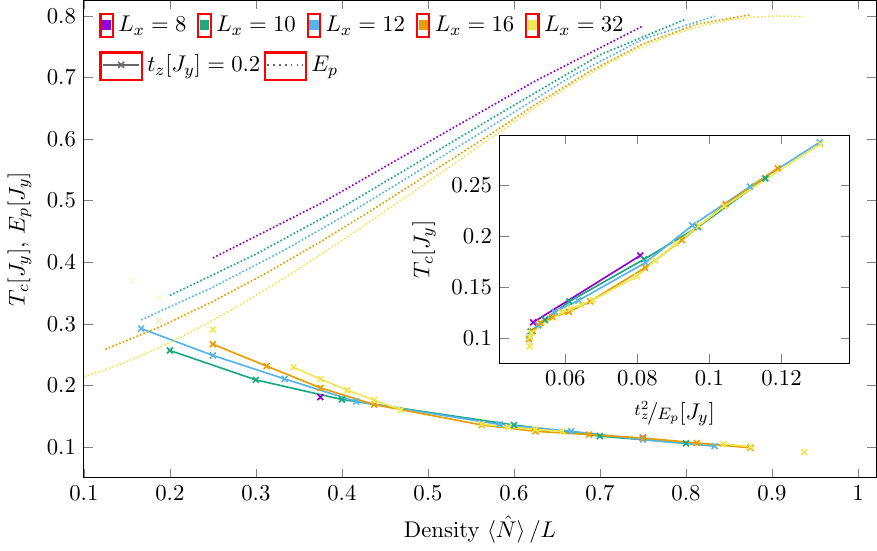}%
	}%
	\caption%
	{%
		\label{fig:Tcs:FiniteSize}%
		Critical temperature $T_c$ vs total density for $t_x=0.7 J_y$, and $t_z=0.2 J_y$ at different $L_x$ (solid lines), computed via \gls{MPSMF}.
		For comparison, $E_p$ from \cref{fig:EpEsCompSos} is shown; $2E_s$ not shown here, as it is consistently above $E_p$.
		We keep to densities for which $T_c\leq \min\{E_p,2E_s\}$ (cf. main text).
		Inset: the same $T_c$\hyp data plotted over $t_z^2/E_p$.
	}%
\end{figure}%
For a pilot modeling of how this 2D \gls{highTc} system would perform when implemented in ultra\hyp cold atomic gases, we focus our \gls{MPSMF} calculations on parameters that have already been realized to a large degree, for a 1D ladder~\cite{Hirthe2023}.
In that original experiment, one had $t_x=0.7 J_y$, $J_x/J_y=0.0476 J_y$ and $L_x=7$.
Here, we neglect $J_x$ and $J_z$ as irrelevant in magnitude, and compute $T_c$ for the onset of superconductivity in the 2D bilayer system from \cref{eq:ratio} and \cref{eq:delta} for $L_x=8$, $10$, $12$, $16$ and $32$.
In order to keep the numerical effort within reason, we are fixing inter\hyp ladder tunneling $t_z$ to be equal to $0.2J_y$.
As the resulting $T_c$\hyp data in \cref{fig:Tcs:FiniteSize} shows, plotted alongside the $E_p$ data from \cref{fig:EpEsCompSos} for reference, this choice of $t_z$ also allows us to explore a wide range of densities:
with $T_c$ increasing with decreasing density, and known to increase with $t_z$~\cite{Bollmark2023}, these \gls{mixD} systems may always reach a point where the predicted $T_c$ matches $E_p$.
Beyond such a point the \gls{MPSMF} framework clearly stops being internally consistent, which is why we perform no calculations when it is clear that this would result in $T_c>E_p$.
Here, with $2E_s>E_p$, that is the key criterion for internal consistency of the \gls{MPSMF} predictions, but in general this criterion naturally must be generalized to be $T_c>\min\{E_p,2E_s\}$
Prioritizing the exploration of a wide range of densities, as we do here, is also sensible because the density remains one of the properties that may still show significant fluctuations from one run to the next even in well-controlled quantum gas microscope set-ups.

There are three key findings in the data shown in \cref{fig:Tcs:FiniteSize}.
First, $T_c$ reaches significant fractions of the spin\hyp exchange coupling $J_y$ across all densities for which we have run \gls{MPSMF} simulations.
These systems, once implemented experimentally, thus absolutely would realize the analogues of repulsively mediated \gls{highTc} superconducting states.
There are three caveats to this statement:
(1) We have not ruled out competing many\hyp body instabilities, which represent one of the biggest challenges for any theory aimed at doped systems of repulsively interacting fermions.
(2) We have not calculated the entropies per particle at these critical temperatures, which are the actual figures of merit when deciding whether a current quantum simulator based on ultra cold atomic gases could actually reach these phases.
(3) Due to the mean\hyp field component of the \gls{MPSMF} framework, these $T_c$\hyp values are inevitably overestimates.
We discuss caveats (1) and (2), which are important and can be resolved, but are outside the scope of the present work, in more detail in \cref{sec:disc}.
Regarding the third caveat, we have been part of previous collaborations that demonstrated that the \gls{MPSMF} framework overestimates $T_c$ by a quasi\hyp constant factor as other parameters of the system are varied~\cite{Bollmark2020a,Bollmark2023}.
This implies that the dependence of $T_c$ on the various microscopic processes is captured correctly, at least at the qualitative level.
With that, the \gls{MPSMF} framework, in which the physics in the direction of dominant coupling are captured quasi\hyp exactly by the \gls{MPS} component, performs far better than standard pure mean\hyp field approaches for fermions.
These pure mean\hyp field techniques typically can only model pair\hyp breaking but not phase\hyp fluctuations, leading to a growing overestimation of $T_c$ as the pairs become heavier at stronger binding~\cite{Tempere2012,Sewer2002}.
With its constant\hyp factor overestimation, the \gls{MPSMF} framework can therefore be used to search, e.g., for the parameters that optimize $T_c$, or for those that optimize the critical entropy per particle.
The one unknown at present is the exact factor by which, e.g., $T_c$ is overestimated.
Due to the nature of the mean\hyp field approximation, this factor will be larger for 2D systems than for 3D ones~\cite{Bollmark2020a,Bollmark2023}.
Here, based on prior work, this factor may be on the order of $4$~\cite{Bollmark2023}.

The second key finding from \cref{fig:Tcs:FiniteSize} is that $T_c$ decreases monotonically as the density is increased.
As $n$ grows, $E_p$ increases monotonically while $E_s$ decreases, and $K_c$ goes from maximum to maximum via its minimum, while $\rho_c$ does the reverse (cf. \cref{fig:EpEsCompSos}).
While all these quantities enter $T_c$ in non\hyp ¸trivial ways, our results show that on balance these 2D bilayer systems put a premium on the constituent t\hyp J ladders simultaneously having a low pairing gap $E_p$ and a high superconducting susceptibility, as parametrized by $K_c$, in order to achieve as high a value of $T_c$ as possible.
With this behavior, and especially the preference for lower $E_p$-values to achieve higher $T_c$, the anisotropic 2D \gls{mixD} system at present parameters behaves analogous to other high-$T_c$ superconducting systems.
Namely, superconductivity in this class is limited by the phase-stiffness of the total 2D or 3D system, as pairs are bound much stronger than in BCS\hyp type superconductor, and are therefore heavy~\cite{emery_importance_1995}. 
Relative to a hypothetical superconductor that is balanced optimally between the conflicting aims of strong pairing and high phase stiffness, this group of superconductors has an ``excess'' of pairing, to the detriment of the phase stiffness.
It is for this reason that shedding $E_p$ by lowering density raises $T_c$.
The phase stiffness in this argument is that of the 2D array.
This is different from the $\rho_c$ of the isolated ladders, being controlled instead by the prefactor $t_z^2/E_p$ of the \gls{pp} amplitudes in \cref{eq:amplitudes}.
As a result, the inset in \cref{fig:Tcs:FiniteSize} shows how $T_c$ in this pilot calculation scales linearly in $t_z^2/E_p$ over most of the parameter\hyp range.
It does not yet enter the superlinear scaling regime that was found in previous work~\cite{Bollmark2023}.

The third key finding is that while the system still shows some finite size effects at lower $t_z^2/E_p$, $T_c$ rapidly stabilizes as this ratio rises, as is also shown in the inset of \cref{fig:Tcs:FiniteSize}.
When plotted versus density, as in the main figure of \cref{fig:Tcs:FiniteSize}, the $T_c$\hyp data collapses on top of each other at higher $n$\hyp values, while exhibiting finite\hyp size effects at lower $n$, where $T_c$ still rises appreciably as $L_x$ grows.

\subsection{Quantifying finite\hyp size effects -- heuristic criterion}\label{sec:heuristic}
The present work aims at modeling possible experiments on current quantum simulators based on ultra cold atomic gases.
At present, that means that one will work at values of $L_x$ that are in the lower double digits.
And while the \gls{MPSMF} pilot calculations in this section show only limited finite\hyp size effects, as discussed in \cref{sec:method}, that will be the case less and less when pushing these systems to higher values of $t_x$ at fixed $L_x$.
When calculating the different relevant properties of the isolated t\hyp J ladders that enter the \gls{MPSMF} framework explicitly or implicitly, it is therefore important to know when the formulas from \gls{TLL} field theory loose their validity due to such finite\hyp size effects.
In our chosen approach, the calculation of the speed of sound $v_c$ of the ladder's symmetric charge mode using \cref{eq:vc} is of particular concern, as it underlies our way of calculating $K_c$ and thus $T_c$.
Specifically, there is no easy way of numerically distinguishing excited states that belong to the symmetric density mode of the ladder, the one with lowest energy being required for \cref{eq:vc}, from states belonging to the gapped antisymmetric density sector~\cite{giamarchi_quantum_2003}.
When the finite\hyp size gaps between the many\hyp body states are smaller than this gap $\Delta_a$, that is no problem.
However, when the finite\hyp size gaps grow larger than $\Delta_a$ with increasing $t_x$ at fixed $L_x$, the constrained \gls{DMRG} calculation that is meant to yield $E_1(0, 2N,L_x)$ for the symmetric density sector in \cref{eq:vc} will instead pick up a state from the antisymmetric sector.

\begin{figure}%
	\centering
	\ifthenelse{\boolean{buildtikzpics}}%
	{%
		\newboolean{clickablePlot}%
		\setboolean{clickablePlot}{false}%
		\input{figures/heuristic_spectrum}%
	}%
	{%
		\includegraphics{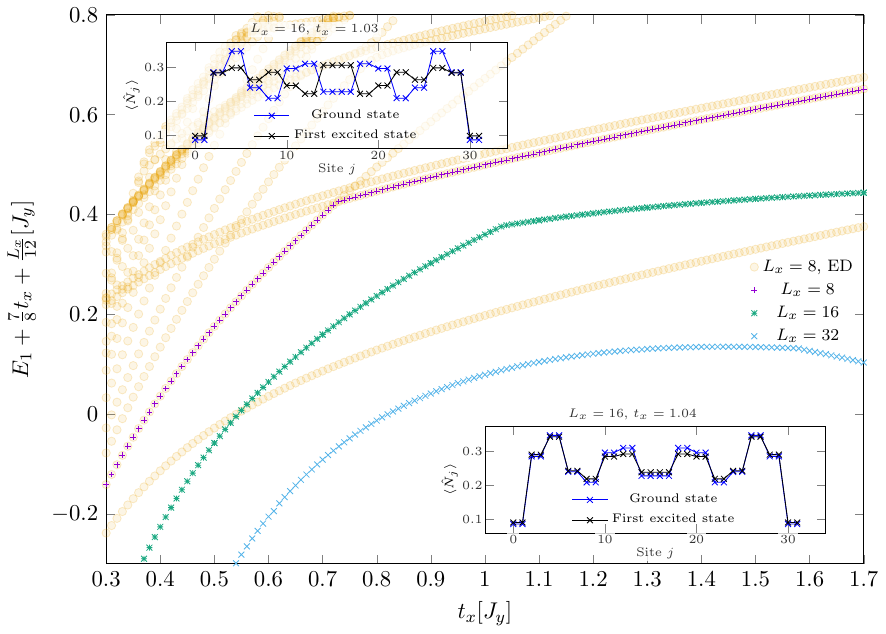}%
	}%
	\caption%
	{%
		\label{fig:heuristicKink}%
		Kink in the energy of the first excited state due to a level\hyp crossing, shown for different system sizes and a density $\frac{\braket{\hat N}}{L}=0.25$.
		The y\hyp axis is tilted to make the kink easier to spot.
		In the inset, the difference in the density distribution is shown left and right of the kink.
		Note that this kink is purely a finite size effect, i.e., the crossover can be prevented by using a large enough system.
	}%
\end{figure}

By comparing \gls{DMRG} calculations with those from \gls{ED}, we have established a heuristic for diagnosing whether the first excited state in the sector $(N,S=0)$ is valid.
In \cref{fig:heuristicKink}, we show the energy of the first excited state for $L_x=8, 16, 32$ and the entire spectrum for $L_x=8$ for a low filling of $N=\nicefrac{L}{2}$, with the tilting function $\frac78 t_x + \frac{L_x}{12}$ being added in order to highlight the effect.
As one can see for the case of $L_x=8$, there is a kink in the energy of the first excited state where a level crossing occurs.
It is this heuristic that underlies the left\hyp out areas of the $t_x$\hyp $n$ parameter\hyp space for $v_c$, $K_c$ and $\rho_c$ in \cref{fig:EpEsCompSos}. 
As also illustrated in \cref{fig:heuristicKink}, the introgression of a state from the antisymmetric density sector with growing $t_x$ announces itself by a sharp change in the profile of the local density:
the profile shifts from that of an excited density distribution relative to that of the absolute ground state at $E_0(0, 2N,L_x)$, to one that looks very similar to it, and which we interpret to actually correspond to that of the ground state of the antisymmetric density sector.
As that graph also illustrates, the point at which this change happens is shifting towards infinite $t_x$ as $L_x$ is raised, as is to be expected.
In this way, we know when to stop increasing $t_x$, in order to avoid having the inputs of the \gls{MPSMF} simulations distorted by finite\hyp size effects.

\section{Discussion}\label{sec:disc}%

The present work prepares the ground for deliberately designing experiments with ultra cold atomic lattice gases to act as quantum simulators for \gls{highTc} superconductivity, as has been predicted for the 2D bilayer \gls{mixD} systems that we model here.
Concretely, we focus on the anisotropic versions of these systems, $t_z < t_x$, as these are amenable to large\hyp scale modeling with the \gls{MPSMF} framework, including the ability to predict how the critical temperature for superconductivity depends on the system's microscopic parameters.
However, due to the \gls{mixD} systems having the proven ability to engineer exceptionally strong pairing energies $E_p$ and spin gaps $E_s$, the \gls{MPSMF} approach could treat these systems much closer to the the isotropic limit, $t_z\approx t_x$, as long as $T_c\leq \min\{E_p,2E_s\}$ remains satisfied.
We plan on addressing these regimes in future work.
We stress that as $t_z$ approaches $\min\{E_p,2E_s\}$, the critical temperature may not necessarily decay -- it may just level off due to the fact that single\hyp fermion tunneling between ladders is now no longer effectively suppressed.
Alternatively, $T_c$ may just start to deviate from the known scaling when $t_z/\min\{E_p,2E_s\}\approx 1$,

With the groundwork provided here, the most demanding challenges for turning 2D \gls{mixD} systems into working simulators of \gls{highTc} superconducting experiments can be taken on in future work.
Two of these challenges already appeared as caveats regarding the calculated $T_c$\hyp values in \cref{subsec:tc}:
how can we rule out that the \gls{highTc} superconducting states that we predict are not actually prevented from forming by competing instabilities, such as insulating phases? 
And what entropies per particle $S/N$ do our calculated $T_c$\hyp values actually correspond to, given that $S/N$ is the actual figure of merit that determines which correlated phases can be reached by atomic lattices gases in practice?
Though computationally demanding, it turns out that the \gls{MPSMF} framework does offer the capabilities to address both these challenges.
Though for simpler systems, we have already been part of a collaboration demonstrating that the \gls{MPSMF} approach can incorporate multiple potential instabilities at the same time~\cite{Bollmark2025}.
With adaptation, these tools can be deployed to the present systems, to see whether insulating or superconducting instabilities order the ground state at any given set of microscopic parameters.
Likewise, the entropy of thermal systems can be computed within the \gls{MPSMF} approach, e.g. via direct calculation of the thermal density matrix using state purification~\cite{schollwock_schollwock_2011}.
As challenging as these calculations are, they will allow to predict to which microscopic parameters an experiment may best be directed in order to maximize the chance of observing \gls{highTc} superconductivity within current limitations on the entropy per particle.
In this way, it will be possible to evaluate, e.g., whether the sweet spot of $E_p$ and $2E_s$ that we have identified in this work would be one such optimal regime.
This will include pushing $T_c$ higher systematically, by increasing $t_z$ at those densities where $E_p$ and $E_s$ would allow for that, which is another angle not yet pursued in the present work that we are going to address in the future.

We point out that the \gls{MPSMF} framework could be refined further to at least partially incorporate the effects of quantum and thermal fluctuations around the mean\hyp field amplitudes \cref{eq:amplitudes}.
Akin to the stability analysis around the saddle points in path integral treatments of superconductivity~\cite{Tempere2012}, the structure of the \gls{MPSMF} approach at finite temperature, as implemented via state purification, can be extended to incorporate small quadratic fluctuations.
This should further reduce the factor by which this theoretical framework overestimates $T_c$. 
In terms of independent validation of the predictions made here, outside of actually performing the experiments, the outlook might be indeterminate at present.
The \gls{MPSMF} framework for fermions was developed because there had been a lack of quantitative numerics that could treat large, anisotropic systems of repulsively interacting itinerant systems in 2D and 3D at low and zero temperature.
One possible method, at least for the 2D bilayer systems studied here, might be phase\hyp free auxiliary\hyp field quantum Monte Carlo~\cite{nguyen_cpmc-lab_2014,qin_coupling_2016,He2019}, should these techniques be amenable to incorporating the constraint on doubly occupied sites at some level.

\paragraph*{Acknowledgments}
A. K. would like to thank Immanuel Bloch, Annabelle Bohrdt, Titus Franz, Thierry Giamarchi, Fabian Grusdt and Si Wang for helpful discussions.
This work was supported by an ERC Starting Grant from the European Union’s Horizon 2020 research and innovation programme under grant agreement No. 758935; and the UK’s Engineering and Physical Sciences Research Council [EPSRC; grant number EP/W022982/1 and UKRI2088]. 
The computations were enabled by resources provided through multiple EPSRC “Access to HPC” calls (Spring 2023, Autumn 2023, Spring 2024 and Autumn 2024, Autumn 2025) on the ARCHER2, Peta4-Skylake and Cirrus compute clusters, as well as by computer time awarded by the National Academic Infrastructure for Supercomputing in Sweden (NAISS). 
This work was supported by a grant from the Swiss National Supercomputing Centre (CSCS) under project ID s1307 on Alps.
We further acknowledge the EuroHPC Joint Undertaking for awarding this project access to the EuroHPC supercomputer LUMI, hosted by CSC (Finland) and the LUMI consortium through an EuroHPC Regular Access call. 
The authors gratefully acknowledge the HPC RIVR consortium and EuroHPC JU for funding this research by providing computing resources of the HPC system Vega at the Institute of Information Science of the Republic of Slovenia. 
The authors also acknowledge the use of the HWU high-performance computing facility (DMOG) and associated support services in the completion of this work.

\appendix

\section{Numerical procedures and tables of results}\label{app:numerics}

For nearly all results in this work we computed ground\hyp state energies and the energy of the first excited state with \gls{DMRG} \cite{white_density_1992,White93,schollwock_schollwock_2011}.
Following common practice, we performed multiple calculations with different bond dimensions, usually at least 
\begin{align}
	\chi = 64, 128, 256, 512, 768, 1024, 1280, 1536, 1792, 2048\;,
\end{align}
in order to extrapolate to zero discarded weight.
Note that generally not every one of these bond dimension resulted in converged results, but we made certain that at least 3 of them did.
The one case where we did not perform extrapolations is the local density profile, which is used in \cref{sec:heuristic}.
There, we used the density profile for bond dimension $\chi=1024$, because the error in the local density is much smaller than the chosen cut\hyp off for the heuristic argument.
While it is possible to use the converged mean\hyp field amplitudes from calculations with lower bond dimensions as starting guesses for those with larger bond dimensions, as discussed in~\cite{Bollmark2023}, we refrained from this in the present work, in order to make as certain as possible  that calculations do not get stuck in local minima.
All shown data, including their fits, are accessible via zenodo on reasonable request \cite{kohler_2026_20737857}.

\section{Analytics miscellanea}\label{app:analytics}
\begin{figure}%
	\centering
	\ifthenelse{\boolean{buildtikzpics}}%
	{%
		\input{figures/R_vs_K_c}%
	}%
	{%
		\includegraphics{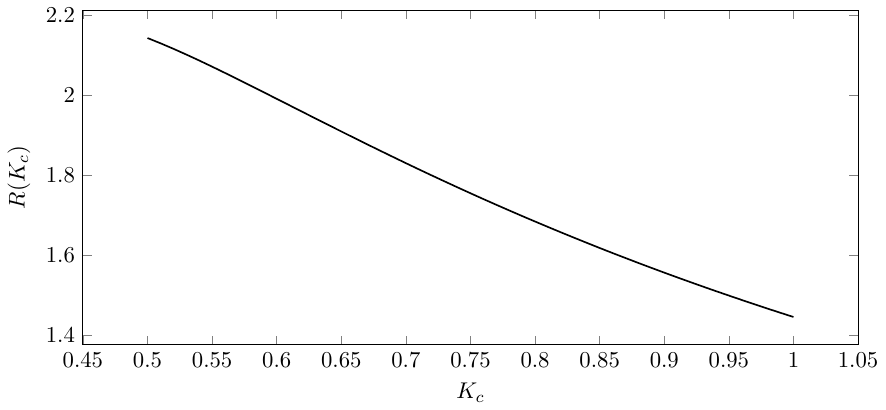}%
	}%
	\caption%
	{%
		\label{fig:R_vs_K_c}
		Ratio function \cref{eq:ratio} for the entire valid range of $K_c$.
	}%
\end{figure}%

As developed in~\cite{Bollmark2023}, the gap between ground and first excited state, \cref{eq:delta}, allows to calculate $T_c$ for the onset of superconductivity directly, \cref{eq:tc}, via the ratio-function $R(K_c)$,~\cref{eq:ratio}.
Within the possible range of $K_c$-values, the ratio-function is shown in \cref{fig:R_vs_K_c}, decreasing monotonically with $K_c$
The function $\kappa(K_c)$ that appears in the definition~\cref{eq:ratio} is given by 
\begin{equation} \label{combination_gamma_functions_ratio}
    \kappa (K_c) = \frac{1}{\pi}\frac{\Gamma \left ( \frac{1}{8K_c}\right ) }{\Gamma \left (1-  \frac{1}{8K_c}\right  )} \left(\frac{\sqrt{\pi}\Gamma \left ( \frac{1}{2} \frac{8K_c}{8K_c-1}\right ) }{2\Gamma \left ( \frac{1}{2}\frac{1}{8K_c-1}\right )} \right)^{2-\frac{1}{4K_c}}\;.
\end{equation}

\begin{bibliography}{literature}
 
\end{bibliography}
\end{document}

%% file: packages.tex
\usepackage{lineno}
\usepackage{pdftexcmds}
\usepackage{graphicx}  
\usepackage{dcolumn}   
\usepackage{bm}        
\usepackage{amssymb}   
\usepackage{amsmath}
\usepackage{blkarray, multirow, graphicx, diagbox, color, colortbl}
\usepackage{bbm, bbold}
\usepackage{ifthen}
\usepackage{booktabs}
\usepackage{xkeyval}
\usepackage{moreverb}
\usepackage{rotating}
\usepackage{slashbox}
\usepackage{xspace}
\usepackage{subcaption}
\usepackage{capt-of}
\newcommand\subfigref[1]{(\protect\subref{#1})}

\usepackage{glossaries}
\glsdisablehyper 
\usepackage{hyphenat}
\usepackage{nicefrac}
\usepackage[]{units}
\usepackage{physics}
\usepackage{braket}
\usepackage{tabto}
\usepackage{listings}
\usepackage{xstring}
\usepackage{gnuplottex}
\usepackage{tikz}
\tikzset{>=stealth}
\usepackage{pgffor}
\usepackage{pgfplots}
\pgfplotsset{compat=newest}
\usepackage{pgfplotstable}
\usepgfplotslibrary{groupplots}
\usetikzlibrary
{
	calc,
	decorations,
	pgfplots.fillbetween,
	pgfplots.patchplots,
	plotmarks,
	patterns,
	positioning,
	petri,
	arrows,
	intersections,
	decorations.markings,
	backgrounds,
	fit,
	matrix,
	graphs,
	shapes.geometric,
	decorations.pathreplacing, 
	decorations.pathmorphing,
	shapes.misc,
	shapes.multipart,
	shapes,
	through,
	tikzmark,
	shadows,
}
\usetikzlibrary{external}
\graphicspath{{figures/autogen/}}
\newboolean{buildtikzpics}
\setboolean{buildtikzpics}{false}

\usepackage[tikz]{ocgx2}

\usepackage{cleveref}  
\Crefname{appendix}{Appendix}{Appendices}
\Crefname{equation}{Equation}{Equations}
\Crefname{figure}{Figure}{Figures}
\Crefname{section}{Section}{Sections}
\Crefname{tabular}{Tabular}{Tabulars}
\crefname{appendix}{App.}{Apps.}
\crefname{equation}{Eq.}{Eqs.}
\crefname{figure}{Fig.}{Figs.}
\crefname{section}{Sec.}{Secs.}
\crefname{tabular}{Tab.}{Tabs.}

\usepackage{xfp}

%% file: acronyms.tex
\newacronym{MPS}{MPS}{matrix\hyp product state}
\newacronym{MPSMF}{MPS+MF}{matrix\hyp product state plus mean\hyp field}
\newacronym{DMRG}{DMRG}{density\hyp matrix renormalization group}
\newacronym{QMC}{QMC}{quantum Monte Carlo}
\newacronym{ED}{ED}{exact diagonalization}
\newacronym{highTc}{high\hyp $T_c$}{high\hyp temperature}
\newacronym{mixD}{mixD}{mixed-dimensional}
\newacronym{NN}{n.n.}{nearest\hyp neighbour}
\newacronym{pp}{pp}{particle\hyp particle}
\newacronym{ph}{ph}{particle\hyp hole}
\newacronym{TLL}{TLL}{Tomonaga\hyp Luttinger liquid}

%% file: defs.tex
\definecolor{mydarkblue}{rgb}{0.0, 0.0, 1.0} 
\definecolor{mydarkgray}{rgb}{0.0, 0.0, 0.0} 
\definecolor{darkyellow}{rgb}{0.5,0.5,0}
\definecolor{darkred}{rgb}{0.5,0.0,0}
\definecolor{darkgreen}{rgb}{0,0.6,0}
\definecolor{slideblue}{rgb}{0.2,0.2,0.7}
\colorlet{primary}{slideblue}
\colorlet{secondary}{slideblue!75!black}
\colorlet{tertiary}{slideblue!50!black}
\colorlet{mylightgray}{black!30!white}
\arrayrulecolor{mydarkblue}
\DeclareCaptionFont{mydarkgray}{\color{mydarkgray}}
\definecolor{lgrey}{gray}{.9}

\let\oldbibliography\thebibliography
\renewcommand{\thebibliography}[1]{%
  \oldbibliography{#1}%
  \setlength{\itemsep}{2pt}%
}

\definecolor{spinupold}{rgb}{.1,.1,.9}   
\colorlet{spinup}{spinupold!50!white}
\definecolor{spindownold}{rgb}{.1,.9,.1} 
\colorlet{spindown}{spindownold!50!white}

\newcommand{\nodagger}[0]{{\phantom{\dagger}}}

\newcommand{\qcsmps}[0]{\textsf{QCS\kern-.45em{\raisebox{-0.1pt}{\footnotesize{\textsf{s}}}}\kern.07em mps}}

%% file: tikzdefs.tex
\usepackage{xstring}
\def\ReplaceStr#1{%
	\IfSubStr{#1}{p}{%
		\StrSubstitute{#1}{p}{.}}{#1}}
\newcommand\ReReplaceStrMinus[1]{%
	\begingroup\expandarg
	\IfSubStr{#1}{-}{%
		\expandafter\StrSubstitute\expandafter{#1}{-}{m}}{#1}
	\endgroup
}
\newcommand\ReReplaceStr[1]{%
	\begingroup\expandarg
	\IfSubStr{#1}{.}{%
		\StrSubstitute{#1}{.}{p}}{#1}
	\endgroup
}

\newcommand\FullReplaceStr[1]{%
	\IfSubStr{#1}{.}
	{
		\StrSubstitute{#1}{.}{p}[#1]
	}{}
	\IfSubStr{#1}{-}
	{
		\StrSubstitute{#1}{-}{m}[#1]
	}{}
}

\makeatletter
\newcommand*{\overlaynumber}{\number\beamer@slideinframe}
\makeatother

\tikzset{
	invisible/.style={opacity=0},
	visible on/.style={alt={#1{}{invisible}}},
	alt/.code args={<#1>#2#3}{%
		\alt<#1>{\pgfkeysalso{#2}}{\pgfkeysalso{#3}} 
	},
}

\newboolean{IncludeHighResFigs}
\setboolean{IncludeHighResFigs}{false}

\ExplSyntaxOn
\DeclareExpandableDocumentCommand \eval { m } { \fp_eval:n { #1 } }
\ExplSyntaxOff

\pgfmathdeclarefunction{linearFct}{2}%
{%
	\pgfmathparse{#1*x+#2}%
}
\pgfmathdeclarefunction{quadFct}{3}%
{%
	\pgfmathparse{#1*x*x+#2*x+#3}%
}
\pgfmathdeclarefunction{logFct}{3}%
{%
	\pgfmathparse{#1*ln(#3*x)+#2}%
}
\pgfmathdeclarefunction{expFct}{2}%
{%
	\pgfmathparse{#1*exp(-x/#2)}%
}

\pgfmathdeclarefunction{Landau}{4}%
{%
	\pgfmathparse{#1/2.0*pow(x,2)+#2/3.0*pow(x,3)+#3/4.0*pow(x,4)+#4/6.0*pow(x,6)}%
}

\tikzset{->-/.style={decoration={
			markings,
			mark=at position .5 with {\arrow{>}}},postaction={decorate}}}

\tikzstyle{orientedsnake} = [
decorate, 
decoration={snake},
->
]  
\tikzstyle{orientedshortarrow} = [
decoration={markings,
	mark=at position .33 with {\arrow{>}}},
postaction={decorate}
]  
\tikzstyle{orientedlongarrow} = [
decoration={markings,
	mark=at position .67 with {\arrow{>}}},
postaction={decorate}
]

\tikzset{dbl/.style={double,
		double equal sign distance,
		-implies,
		shorten >=10pt,
		shorten <=10pt}}

\tikzset{
	between/.style args={#1 and #2}{
		at = ($(#1)!0.5!(#2)$)
	}
}

\definecolor{colorA}{rgb} {0.58,0,0.8275}
\definecolor{colorB}{rgb} {0.11,0.663,0.51}
\definecolor{colorC}{rgb} {0.3373,0.7059,0.9137}
\definecolor{colorD}{rgb} {0.902,0.6235,0}
\definecolor{colorE}{rgb} {0.9451,0.902,0.3255}
\definecolor{colorF}{rgb} {0.3373,0.3255,0.902}
\definecolor{colorG}{rgb} {0.9451,0.3255,0.3373}
\definecolor{colorIPphase}{rgb} {0.4, 0.7, 0.4}
\definecolor{colorIPcharge}{rgb} {0.7, 0.1, 0.7}

\def\gpmarkers{{"+","x","star","square","square*","o","*","triangle","triangle*"}}
\def\gpcolors{{"colorA","colorB","colorC","colorD","colorE"}}

\newboolean{simpletrue}
\newboolean{simplefalse}
\setboolean{simpletrue}{true}
\setboolean{simplefalse}{false}

\tikzset{
  -|-/.style={
    to path={
      (\tikztostart) -| ($(\tikztostart)!#1!(\tikztotarget)$) |- (\tikztotarget)
      \tikztonodes
    }
  },
  -|-/.default=0.5,
  |-|/.style={
    to path={
      (\tikztostart) |- ($(\tikztostart)!#1!(\tikztotarget)$) -| (\tikztotarget)
      \tikztonodes
    }
  },
  |-|/.default=0.5,
}

\pgfmathdeclarefunction{FctPlus}{1}%
{%
	\pgfmathparse{1.0/#1*(x-21.5)}%
}
\pgfmathdeclarefunction{FctMinus}{1}%
{%
	\pgfmathparse{1.0/#1*(-x+21.5)}%
}
\pgfmathdeclarefunction{FctPart}{2}%
{%
	\pgfmathparse{1.0/#1*(-x+21.5)+#2}%
}

\newcommand{\SystemsInternal}[9]
{
	\pgfmathparse{int(#1-1)}%
	\pgfmathsetmacro{\xmax}{\pgfmathresult}%
	\pgfmathparse{int((#2*#6-1))}%
	\pgfmathsetmacro{\ymax}{\pgfmathresult}%
	\pgfmathparse{int((#3*#7-1))}%
	\pgfmathsetmacro{\zmax}{\pgfmathresult}%
	\foreach \z in {0,...,\zmax}%
	{%
		\foreach \x in {0,...,\xmax}%
		{%
			\foreach \y in {0,...,\ymax}%
			{%
				\ifthenelse{\y>0}
				{%
					\node[circle, draw = yellow!55!black, inner sep = 0pt, outer sep = 0pt] at ($(\x*1.25*#4 - \z*1.0*#4, \y*1.25*#4 - \z*0.5*#4)$) (\x\y\z){\pgfmathparse{int(5*#4)}\fontsize{\pgfmathresult pt}{10pt}#5};%
				}%
				{%
					\node[circle, draw = red!55!black, inner sep = 0pt, outer sep = 0pt] at ($(\x*1.25*#4 - \z*1.0*#4, \y*1.25*#4 - \z*0.5*#4)$) (\x\y\z){\pgfmathparse{int(5*#4)}\fontsize{\pgfmathresult pt}{10pt}#5};%
				}%
			}%
		}%
	}%
	\foreach \z [remember=\z as \lastz (initially 0)] in {0,...,\zmax}%
	{%
		\foreach \x [remember=\x as \lastx (initially 0)] in {0,...,\xmax}%
		{%
			\foreach \y [remember=\y as \lasty (initially 1)] in {0,...,\ymax}%
			{%
				\ifthenelse{\y>0}%
				{%
					\pgfmathparse{int(mod(\y,#2))}%
					\ifthenelse{\pgfmathresult=0}%
					{%
						\draw[densely dotted] (\x\lasty\z) -- node[left, inner ysep = 0.5em] {} (\x\y\z);%
					}%
					{%
						\draw[thick] (\x\lasty\z) -- node[left, inner ysep = 0.5em] {} (\x\y\z);%
					}%
				}{}%
				\ifthenelse{\x>0}%
				{%
					\pgfmathparse{int(mod(\x,#2))}%
					\ifthenelse{\pgfmathresult=0}%
					{%
						\begin{scope}[on background layer]
							\draw[opacity=#8, thick] (\lastx\y\z.east) -- node[above, inner xsep = 0.5em, inner ysep = 0.1em] {} (\x\y\z.west);%
						\end{scope}
					}%
					{%
						\draw[thick] (\lastx\y\z) -- node[above, inner xsep = 0.5em, inner ysep = 0.1em] {} (\x\y\z);%
					}%
				}{}%
				\draw[densely dotted] (\x\y\lastz) -- (\x\y\z);
			}%
		}%
	}%
}
\newcommand{\Systems}[9]%
{%
	\tikzset{external/export next=false}%
	\begin{tikzpicture}[baseline]%
		\SystemsInternal{#1}{#2}{#3}{#4}{#5}{#6}{#7}{#8}{#9}
	\end{tikzpicture}%
}%
\newcommand{\Fermions}[5]%
{%
	\pgfmathparse{rnd}%
	\pgfmathsetmacro{\uprnd}{\pgfmathresult}%
	\ifthenelse{\lengthtest{\uprnd pt > #1pt }}%
	{%
		{\color{#2}$\uparrow$}%
	}%
	{%
		\color{#3}$\uparrow$%
	}%
	\pgfmathparse{rnd}%
	\pgfmathsetmacro{\downrnd}{\pgfmathresult}%
	\ifthenelse{\lengthtest{\downrnd pt > #1pt }}%
	{%
		{\color{#4}$\downarrow$}%
	}%
	{%
		{\color{#5}$\downarrow$}%
	}%
}%
\newcommand{\tJs}[4]%
{%
	\pgfmathparse{rnd}%
	\pgfmathsetmacro{\particlernd}{\pgfmathresult}%
	\ifthenelse{\lengthtest{\particlernd pt > #1pt }}%
	{%
		\pgfmathparse{rnd}%
		\pgfmathsetmacro{\spinrnd}{\pgfmathresult}%
		\ifthenelse{\lengthtest{\spinrnd pt > 0.5pt }}%
		{%
			{\color{#2}$\uparrow$}%
		}%
		{%
			{\color{#3}$\downarrow$}%
		}%
	}%
	{%
		{$\vphantom{\uparrow}$\color{#4}\textbullet}%
	}%
}%
\newcommand{\Bosons}[3]%
{%
	\pgfmathparse{rnd}%
	\pgfmathsetmacro{\rnd}{\pgfmathresult}%
	\ifthenelse{\lengthtest{\rnd pt > #1pt }}%
	{%
		{\color{#2}\textbullet}%
	}%
	{%
		{\color{#3}\textbullet}%
	}%
}%

\pgfmathdeclarefunction{lg10}{1}{%
	\pgfmathparse{ln(#1)/ln(10)}%
}

\tikzstyle{PaPcalc} = [draw, ellipse, outer sep=0pt,inner sep=5pt,minimum width=20em, drop shadow={Blue}, fill=SkyBlue, align=center]

\tikzstyle{PaPres} = [draw, outer sep=0pt,inner sep=10pt,minimum width=20em, drop shadow={Blue}, fill=SkyBlue, align=center]

\makeatletter
\def\pgfplotsutil@decstringcounter#1{%
	\begingroup
	\c@pgf@counta=#1\relax
	\advance\c@pgf@counta by -1
	\edef#1{\the\c@pgf@counta}%
	\pgfmath@smuggleone#1%
	\endgroup
}%

\pgfplotsset{
	/pgfplots/each nth point*/.style 2 args={%
		/pgfplots/x filter/.append code={%
			\ifnum\coordindex=0
			\def\c@pgfplots@eachnthpoint@xfilter{0}%
			\edef\c@pgfplots@eachnthpoint@xfilter@cmp{#1}%
			\else
			\ifnum\coordindex>#2\relax
			\pgfplotsutil@advancestringcounter\c@pgfplots@eachnthpoint@xfilter
			\ifx\c@pgfplots@eachnthpoint@xfilter@cmp\c@pgfplots@eachnthpoint@xfilter
			\def\c@pgfplots@eachnthpoint@xfilter{0}%
			\else
			\let\pgfmathresult\pgfutil@empty
			\fi
			\fi
			\fi
		}%
	},
}
\makeatother

%% file: figures/PaP.tex
\tikzsetnextfilename{PaP}
\begin{tikzpicture}
	\node[PaPcalc, minimum width=1em] (gs) {\textbf{ground states wrt $\hat H_{\mathrm{tJ-ladder}}$} \\[0.25em] $\ket{E_0} \forall N, S \in \{0, \nicefrac12, 1\}$};
	\node[PaPres, below left = of gs, xshift=8em, minimum width=1em] (gapsFromGs) {\textbf{pairing energy} \\[0.25em] $E_p = 2 E_0(N+1,S+\nicefrac12)$ \\ $\phantom{E_p=123456} - E_0(N,S) - E_0(N+2,S)$\\[0.5em] \textbf{spin gap} \\[0.25em] $E_s = E_0(N,S+1) - E_0(N,S)$\\[0.5em] \textbf{compressibility $\kappa$} \\[0.25em] $\frac{1}{\kappa} =  \frac{L_x}{4} \left[E_0(N+2,S) + E_0(N-2,S)\right.$\\$\left. - 2E_0(N,S)\right]\phantom{01234567}$};
	\node[PaPcalc, below right = of gs, xshift=-8em, minimum width=1em] (exc) {\textbf{first excited states}\\\textbf{ wrt $\hat H_{\mathrm{tJ-ladder}}$} \\[0.25em] $\ket{E_1} \forall N, S =0$};
	\node[PaPres, below = of exc, minimum width=1em] (sos) {\textbf{speed of sound} \\[0.25em] $v_c = \frac{L_x}{\pi} \left[E_1(N,S) - E_0(N,S)\right]$};
	\node[PaPres, below = of sos, minimum width=1em] (krho) {\textbf{Luttinger parameter} \\[0.25em] $K_c = \frac{\pi}{4} v_c \kappa $};
	\node[PaPcalc, below = of gapsFromGs, minimum width=1em] (scl) {\textbf{self\hyp consistent} \\ \textbf{ground states wrt $\hat H_{\mathrm{effective}}$} \\[0.25em] $\ket{E_0, \vec \alpha} \forall N, S=0$};
	\node[PaPcalc, below = of krho, minimum width=1em, xshift=-1em] (excmpsmf) {\textbf{first excited states}\\\textbf{ wrt $\hat H_{\mathrm{effective}}$} \\[0.25em] $\ket{E_1, \vec \alpha} \forall N, S =0$ \\[0.25em] for converged $\vec \alpha$};
	\node[PaPres, below = of scl, minimum width=1em] (gapFromMPSMF) {\textbf{gap between $\ket{E_0, \vec \alpha}$ and $\ket{E_1, \vec \alpha}$} \\[0.25] $\Delta_\rho = E_1(N,S,\vec \alpha) - E_0(N,S,\vec \alpha)$};
	\node[PaPres, below = of gapFromMPSMF, xshift=12em, minimum width=1em] (Tc) {\textbf{critical temperature} via ratio\footnote{$R(K_c)=2 \pi \left [ \frac{K_c \tan{\left ( \frac{\pi}{2} \frac{1}{4K_c-1}\right )} }{ 2 \kappa^2(K_c/2)(4K_c - 1) \sin \left ( \frac{\pi}{2K_c} \right )  B^2 \left (\frac{1}{4K_c}, 1 - \frac{1}{2K_c} \right ) }  \right ]^{\frac{K_c}{2K_c-1}} \sin \left ( \frac{\pi}{4K_c-1}\right )$} \\[0.25] $T_c= \frac{\Delta_\rho}{R(K_c)}$};
	
	\draw[->, ultra thick] (gs) -- (gapsFromGs);
	\draw[->, ultra thick] (gs) to (exc);
	\draw[->, ultra thick] (exc) -- (sos);
	\draw[->, ultra thick] (sos) -- (krho);
	\draw[->, ultra thick] (gapsFromGs) to (krho.west);
	\draw[->, ultra thick] (scl) -- (excmpsmf);
	\draw[->, ultra thick] (excmpsmf) -- (gapFromMPSMF);
	\draw[->, ultra thick] (scl) to (gapFromMPSMF);
	\draw[->, ultra thick] (gapFromMPSMF) -- (Tc);
	\draw[->, ultra thick, bend left=60, looseness=1.6] (krho) to (Tc);
	\draw[->, ultra thick] 	(gapsFromGs) to (scl);
\end{tikzpicture}

%% file: figures/EpEsCompSos.tex
\tikzsetnextfilename{EpEsCompSos}%
\def\Lxs{{8,10,12,16,32,48,64}}%
\pgfmathtruncatemacro{\LxsDim}{dim(\Lxs)-1}%
\begin{tikzpicture}%
	\begin{groupplot}%
	[%
		group style 	= 	{%
			group size 			=	2 by 3,%
			horizontal sep		=	1em,%
			vertical sep		=	1em,%
			x descriptions at	=	edge bottom,%
			y descriptions at	=	all,%
		},%
		width = 0.5\textwidth-1.8pt,%
		height = 0.225\textheight,%
		xlabel = {Density $\braket{\hat N}/L$},%
		legend pos = north west,%
		legend columns=2,%
		legend style = {draw=none, fill=none},%
		ymin = 0,%
		ymax = 1.1,%
	]%
		\nextgroupplot
		[
			ylabel = {Pairing Energy $E_p[J_y]$},%
		]
			\foreach \i in {0,...,\LxsDim}%
			{%
				\pgfmathsetmacro{\marker}{\gpmarkers[mod(\i,dim(\gpmarkers))]}%
				\pgfmathsetmacro{\cl}{\gpcolors[mod(\i,dim(\gpcolors))]}
				\pgfmathsetmacro{\Lx}{\Lxs[\i]}%
				\edef\plot%
				{%
					\noexpand\addplot%
					[%
						mark	=	\marker,%
						color	=	\cl,%
						thick,%
						error bars/.cd,%
						y dir=both,%
						y explicit,%
					]%
						table%
						[%
							x expr = \noexpand\thisrowno{0}/(2.0*\Lx),%
							y expr = \noexpand\thisrowno{2},%
							y error expr = \noexpand\thisrowno{3},%
						]%
							{../data/Ly_2/Lx_\Lx/pairing_energies_tx_0p7.dat};%
				}\plot%
			}%
			\node[anchor = north west] at (axis cs: 0,1) {$t_x=0.7$};%
		\nextgroupplot%
		[%
			yticklabel pos	=	right,%
			ylabel 			=	{Spin gap $E_s[J_y]$},%
		]%
			\foreach \i in {0,...,\LxsDim}%
			{%
				\pgfmathsetmacro{\marker}{\gpmarkers[mod(\i,dim(\gpmarkers))]}%
				\pgfmathsetmacro{\cl}{\gpcolors[mod(\i,dim(\gpcolors))]}
				\pgfmathsetmacro{\Lx}{\Lxs[\i]}%
				\edef\plot%
				{%
					\noexpand\addplot%
					[%
						mark	=	\marker,%
						color	=	\cl,%
						thick,%
						error bars/.cd,%
						y dir=both,%
						y explicit,%
					]%
						table%
						[%
							x expr = \noexpand\thisrowno{0}/(2.0*\Lx),%
							y expr = \noexpand\thisrowno{2},%
							y error expr = \noexpand\thisrowno{3},%
						]%
							{../data/Ly_2/Lx_\Lx/spin_gaps_tx_0p7.dat};%
					\noexpand\addlegendentry{\noexpand\small $L_x=\Lx$};%
				}\plot%
			}%
		\nextgroupplot%
		[%
			ylabel	=	{Incompressibility $\frac{1}{\kappa}[J_y]$},%
			ymax	=	2,%
		]%
			\foreach \i in {0,...,\LxsDim}%
			{%
				\pgfmathsetmacro{\marker}{\gpmarkers[mod(\i,dim(\gpmarkers))]}%
				\pgfmathsetmacro{\cl}{\gpcolors[mod(\i,dim(\gpcolors))]}
				\pgfmathsetmacro{\Lx}{\Lxs[\i]}%
				\edef\plot%
				{%
					\noexpand\addplot%
					[%
						mark	=	\marker,%
						color	=	\cl,%
						thick,%
						error bars/.cd,%
						y dir=both,%
						y explicit,%
					]%
						table%
						[%
							x expr = \noexpand\thisrowno{0}/(2.0*\Lx),%
							y expr = \noexpand\thisrowno{2},%
							y error expr = \noexpand\thisrowno{3},%
						]%
							{../data/Ly_2/Lx_\Lx/compressibilities_tx_0p7.dat};%
				}\plot%
			}%
		\nextgroupplot%
		[%
			yticklabel pos	=	right,%
			ylabel 			=	{Speed of sound $v_c[J_y]$},%
			ymax			=	2,%
		]%
			\foreach \i in {0,...,\LxsDim}%
			{%
				\pgfmathsetmacro{\marker}{\gpmarkers[mod(\i,dim(\gpmarkers))]}%
				\pgfmathsetmacro{\cl}{\gpcolors[mod(\i,dim(\gpcolors))]}
				\pgfmathsetmacro{\Lx}{\Lxs[\i]}%
				\edef\plot%
				{%
					\noexpand\addplot%
					[%
						mark	=	\marker,%
						color	=	\cl,%
						thick,%
						error bars/.cd,%
						y dir=both,%
						y explicit,%
					]%
						table%
						[%
							x expr = \noexpand\thisrowno{0}/(2.0*\Lx),%
							y expr = \noexpand\thisrowno{2},%
							y error expr = \noexpand\thisrowno{3},%
						]%
							{../data/Ly_2/Lx_\Lx/speed_of_sounds_tx_0p7.dat};%
				}\plot%
			}%
		\nextgroupplot%
		[%
			ylabel	=	{$K_c$},%
			ymax	=	1,%
			ymin	=	0,%
		]%
			\foreach \i in {0,...,\LxsDim}%
			{%
				\pgfmathsetmacro{\marker}{\gpmarkers[mod(\i,dim(\gpmarkers))]}%
				\pgfmathsetmacro{\cl}{\gpcolors[mod(\i,dim(\gpcolors))]}
				\pgfmathsetmacro{\Lx}{\Lxs[\i]}%
				\edef\plot%
				{%
					\noexpand\addplot%
					[%
						mark	=	\marker,%
						color	=	\cl,%
						thick,%
						error bars/.cd,%
						y dir=both,%
						y explicit,%
					]%
						table%
						[%
							x expr = \noexpand\thisrowno{0}/(2.0*\Lx),%
							y expr = 0.25*3.14159265358979323846264*\noexpand\thisrowno{5}/\noexpand\thisrowno{2},%
							y error expr = \noexpand\thisrowno{3},%
						]%
							{../data/results/ratio_Lx_\Lx_tx_0p7.dat};%
				}\plot%
			}%
		\nextgroupplot%
		[%
			yticklabel pos	=	right,%
			ylabel 			=	{$\rho_c[J_y]$},%
			ymax			=	0.25,%
		]%
			\foreach \i in {0,...,\LxsDim}%
			{%
				\pgfmathsetmacro{\marker}{\gpmarkers[mod(\i,dim(\gpmarkers))]}%
				\pgfmathsetmacro{\cl}{\gpcolors[mod(\i,dim(\gpcolors))]}
				\pgfmathsetmacro{\Lx}{\Lxs[\i]}%
				\edef\plot%
				{%
					\noexpand\addplot%
					[%
						mark	=	\marker,%
						color	=	\cl,%
						thick,%
						error bars/.cd,%
						y dir=both,%
						y explicit,%
					]%
						table%
						[%
							x expr = \noexpand\thisrowno{0}/(2.0*\Lx),%
							y expr = \noexpand\thisrowno{5}*\noexpand\thisrowno{5}/(4*\noexpand\thisrowno{2}),%
							y error expr = \noexpand\thisrowno{3},%
						]%
							{../data/results/ratio_Lx_\Lx_tx_0p7.dat};%
				}\plot%
			}%
	\end{groupplot}%
\end{tikzpicture}%

%% file: figures/pairingenergyANDspingapSurfacePlot.tex
		\tikzsetnextfilename{pairingenergyANDspingapSurfacePlot}
		\begin{tikzpicture}
			\pgfplotsset{colormap={PairingSpingap}{color=(colorA) color=(colorB)}}
			\begin{axis}
			[
				width		= 0.95\textwidth-3.7pt,                
				height		= 0.4\textheight, 
				view		= {30}{10}, 
				xlabel		= {$t_x$}, 
				ylabel		= {$N$}, 
				zlabel		= {\textcolor{colorA}{pairing energy $E_p[J_y]$}, \textcolor{colorB}{spin gap $2\cdot E_s[J_y]$}},
				zlabel style= {xshift = -1em},
				zmin		= 0,
				z buffer	= sort,
			]
				\addplot3
				[
					patch,
					patch type 					= rectangle,
					patch table with point meta	= {data/patch_table_EpEsAll.dat},
					shader						= faceted interp,
					patch refines				= 2,
					restrict x to domain		= 0:5,
					opacity						= 0.6,
				]
					table
					[
						x expr	= {\thisrowno{0}},
						y expr	= {\thisrowno{1}},
						z expr	= {\thisrowno{3}},
					]
						{data/EpEsAll.dat};
				\coordinate (inset) at (-0.0,0,1.925);
			\end{axis}
			\node
			[
				inner sep = 0pt,
				anchor = north west,
				at = {(inset)},
				fill=white,
				fill opacity = 0.9,
				minimum width=0.3\textwidth-2.2em,,
				minimum height = 0.2\textwidth-2.6em,%
			]{};
			\begin{axis}
			[
				anchor = north west,
				at = {(inset)},
				width=0.3\textwidth,
				height = 0.2\textwidth,%
				ymin=0.5,
				ymax=1.7,
				xlabel = {\footnotesize $t_x^{\nicefrac{1}{3}}$},
				ylabel = {\footnotesize \textcolor{colorA}{$E_p[J_y]$}},
				xlabel shift=-0.5em,
				ylabel shift=-0.5em,
				axis y line*=left,%
				title={\footnotesize $L_x=16$, $N=30$},
				title style= {yshift=-0.75em},
				every tick label/.append style={font=\footnotesize},
			]
				\addplot
				[
					mark = o,
					thick,
					colorA
				]
					table
					[
						x expr = {\thisrowno{0}^(1/3)},
						y expr = {\thisrowno{3}},
					]
						{data/pairing_energies_Lx_16_N_max.dat};
			\end{axis}
			\begin{axis}
			[
				anchor = north west,
				at = {(inset)},
				width=0.3\textwidth,
				height = 0.2\textwidth,%
				ymin=0.25,
				ymax=0.85,
				ylabel = {\footnotesize\textcolor{colorB}{$E_s[J_y]$}},
				ylabel style={fill = white, fill opacity = 0.9},
				axis y line*=right,%
				axis x line=none,%
				every tick label/.append style={font=\footnotesize},
			]
				\addplot
				[
					mark = x,
					thick,
					colorB
				]
					table
					[
						x expr = {\thisrowno{0}^(1/3)},
						y expr = {\thisrowno{3}},
					]
						{data/spin_gaps_Lx_16_N_max.dat};
			\end{axis}
		\end{tikzpicture}

%% file: figures/Incompressibility_SoS_K_rho_Lx_32.tex
\tikzsetnextfilename{Incompressibility_SoS_K_rho_Lx_32}%
\begin{tikzpicture}
	\begin{groupplot}
	[
		group style 	= 	{%
			group size 			=	2 by 2,%
			horizontal sep		=	3em,%
			vertical sep		=	1em,%
			x descriptions at	=	edge bottom,%
			y descriptions at	=	edge left,%
		},%
		width=0.5\textwidth-4pt,
		height=0.35\textwidth,
		xlabel={Density},
		ylabel={$t_x[J_y]$},
		xmin={3/64},
		xmax={61/64},
		ymin={0.3},
		ymax={5.5},
	]
		\nextgroupplot
		[
			colorbar,
			colormap/viridis,
			enlargelimits=false,
			axis on top,
			colorbar style=%
			{%
				title = {$\nicefrac{1}{\kappa}[J_y]$},
				title style = {yshift=-0.5em, xshift=0.6em},
				width=0.5em,
				height=0.2\textwidth,
				anchor=south west,
				at={(1.04,0)},
			},%
		]
			\addplot
			[
				matrix plot*,
				mesh/cols=29,
				mesh/rows=13,
				point meta=explicit,
				shader=flat
			] 
				table 
				[
					x expr = \thisrowno{1}/64,
					y expr = \thisrowno{0},
					meta index = 3
				] 
					{data/results/ratio_Lx_32.dat};
					
		\nextgroupplot
		[
			colorbar,
			colormap/viridis,
			enlargelimits=false,
			axis on top,
			colorbar style=%
			{%
				title = {$v_c[J_y]$},
				title style = {yshift=-0.5em, xshift=0.6em},
				width=0.5em,
				height=0.2\textwidth,
				anchor=south west,
				at={(1.04,0)},
			},%
			xmin={3/64},
		]
			\addplot
			[
				matrix plot*,
				mesh/cols=29,
				mesh/rows=13,
				point meta=explicit,
				shader=flat
			] 
				table 
				[
					x expr = \thisrowno{1}/64,
					y expr = \thisrowno{0},
					meta index = 6
				] 
					{data/results/ratio_Lx_32.dat};
		\nextgroupplot
		[
			colorbar,
			colormap/viridis,
			enlargelimits=false,
			axis on top,
			colorbar style=%
			{%
				title = {$K_c$},
				title style = {yshift=-0.5em, xshift=0.6em},
				width=0.5em,
				height=0.2\textwidth,
				anchor=south west,
				at={(1.04,0)},
			},%
		]
			\addplot
			[
				matrix plot*,
				mesh/cols=29,
				mesh/rows=13,
				point meta=explicit,
				shader=flat
			] 
				table 
				[
					x expr = \thisrowno{1}/64,
					y expr = \thisrowno{0},
					meta expr = 0.25*3.14159265358979323846264*\thisrowno{6}/\thisrowno{3}
				] 
					{data/results/ratio_Lx_32.dat};
		\nextgroupplot
		[
			colorbar,
			colormap/viridis,
			enlargelimits=false,
			axis on top,
			colorbar style=%
			{%
				title = {$\rho_c[J_y]$},
				title style = {yshift=-0.5em, xshift=0.6em},
				width=0.5em,
				height=0.2\textwidth,
				anchor=south west,
				at={(1.04,0)},
			},%
		]
			\addplot
			[
				matrix plot*,
				mesh/cols=29,
				mesh/rows=13,
				point meta=explicit,
				shader=flat
			] 
				table 
				[
					x expr = \thisrowno{1}/64,
					y expr = \thisrowno{0},
					meta expr = \thisrowno{6}*\thisrowno{6}/(4*\thisrowno{3}),
				] 
					{data/results/ratio_Lx_32.dat};
	\end{groupplot}
	\begin{groupplot}
	[
		group style 	= 	{%
			group size 			=	2 by 2,%
			horizontal sep		=	3em,%
			vertical sep		=	1em,%
			x descriptions at	=	edge bottom,%
			y descriptions at	=	edge left,%
		},%
		width=0.5\textwidth-4pt,
		height=0.35\textwidth,
		xlabel={Density},
		ylabel={$t_x[J_y]$},
		xmin={3/64},
		xmax={61/64},
		ymin={0.3},
		ymax={5.5},
	]
		\nextgroupplot
		[
			colorbar,
			colormap/viridis,
			enlargelimits=false,
			axis on top,
			colorbar style=%
			{%
				title = {$\nicefrac{1}{\kappa}[J_y]$},
				title style = {yshift=-0.5em, xshift=0.6em},
				width=0.5em,
				height=0.2\textwidth,
				anchor=south west,
				at={(1.04,0)},
			},%
		]
			\addplot
			[
				matrix plot*,
				mesh/cols=29,
				mesh/rows=13,
				point meta=explicit,
				shader=flat
			] 
				table 
				[
					x expr = \thisrowno{1}/64,
					y expr = \thisrowno{0},
					meta index = 3
				] 
					{data/results/ratio_Lx_32.dat};
					
		\nextgroupplot
		[
			colorbar,
			colormap/cool,
			enlargelimits=false,
			axis on top,
			colorbar style=%
			{%
				title = {$v_c[J_y]$},
				title style = {yshift=-0.5em, xshift=0.6em},
				width=0.5em,
				height=0.2\textwidth,
				anchor=south west,
				at={(1.04,0)},
				opacity=0,
			},%
		]
			\addplot
			[
				matrix plot*,
				mesh/cols=29,
				mesh/rows=13,
				point meta=explicit,
				shader=flat,
				opacity=0,
			] 
				table 
				[
					x expr = \thisrowno{1}/64,
					y expr = \thisrowno{0},
					meta index = 6
				] 
					{data/results/ratio_Lx_32.dat};
			\addplot
			[
				matrix plot*,
				mesh/cols=31,
				mesh/rows=13,
				point meta=explicit,
				shader=flat,
				opacity=0.9
			] 
				table 
				[
					x expr = \thisrowno{0}/64,
					y expr = \thisrowno{1},
					meta expr = {\thisrowno{2}>exp(-x/(0.5)-2) ? nan : 0}
				] 
					{data/density_profile_difference_Lx_32.dat};
		\nextgroupplot
		[
			colorbar,
			colormap/cool,
			enlargelimits=false,
			axis on top,
			colorbar style=%
			{%
				title = {$K_c$},
				title style = {yshift=-0.5em, xshift=0.6em},
				width=0.5em,
				height=0.2\textwidth,
				anchor=south west,
				at={(1.04,0)},
				opacity=0
			},%
		]
			\addplot
			[
				matrix plot*,
				mesh/cols=29,
				mesh/rows=13,
				point meta=explicit,
				shader=flat,
				opacity=0,
			] 
				table 
				[
					x expr = \thisrowno{1}/64,
					y expr = \thisrowno{0},
					meta expr = 0.25*3.14159265358979323846264*\thisrowno{6}/\thisrowno{3}
				] 
					{data/results/ratio_Lx_32.dat};
			\addplot
			[
				matrix plot*,
				mesh/cols=31,
				mesh/rows=13,
				point meta=explicit,
				shader=flat,
				opacity=0.9
			] 
				table 
				[
					x expr = \thisrowno{0}/64,
					y expr = \thisrowno{1},
					meta expr = {\thisrowno{2}>exp(-x/(0.5)-2) ? nan : 0}
				] 
					{data/density_profile_difference_Lx_32.dat};
		\nextgroupplot
		[
			colorbar,
			colormap/cool,
			enlargelimits=false,
			axis on top,
			colorbar style=%
			{%
				title = {$\rho_c[J_y]$},
				title style = {yshift=-0.5em, xshift=0.6em},
				width=0.5em,
				height=0.2\textwidth,
				anchor=south west,
				at={(1.04,0)},
				opacity=0
			},%
			xmin={3/64},
		]
			\addplot
			[
				matrix plot*,
				mesh/cols=29,
				mesh/rows=13,
				point meta=explicit,
				shader=flat,
				opacity=0,
			] 
				table 
				[
					x expr = \thisrowno{1}/64,
					y expr = \thisrowno{0},
					meta expr = \thisrowno{6}*\thisrowno{6}/(4*\thisrowno{3}),
				] 
					{data/results/ratio_Lx_32.dat};
			\addplot
			[
				matrix plot*,
				mesh/cols=31,
				mesh/rows=13,
				point meta=explicit,
				shader=flat,
				opacity=0.9
			] 
				table 
				[
					x expr = \thisrowno{0}/64,
					y expr = \thisrowno{1},
					meta expr = {\thisrowno{2}>exp(-x/(0.5)-2) ? nan : 0}
				] 
					{data/density_profile_difference_Lx_32.dat};
	\end{groupplot}
\end{tikzpicture}

%% file: figures/all_Tcs_with_Ep.tex
\tikzsetnextfilename{all_Tc_with_Ep}%
\begin{tikzpicture}%
	\def\tperpD{0.2}
	\def\tperpS{0p2}
	\def\marker{x}
	\def\Lxs{{8,10,12,16,32}}%
	\pgfmathtruncatemacro{\LxsDim}{dim(\Lxs)-1}%
	\begin{axis}%
	[%
		width = 1.0\textwidth,%
		height = 0.4\textheight,%
		xlabel = {Density $\braket{\hat N}/L$},%
		ylabel = {$T_c[J_y]$, $E_p[J_y]$},%
		ymin	=	0.05,
		ymax	=	0.825,
		xmin	=	0.1,
	]%
		\foreach \i in {0,...,\LxsDim}%
		{%
			\pgfmathsetmacro{\cl}{\gpcolors[mod(\i,dim(\gpcolors))]}
			\edef\colorlegend%
			{%
				\noexpand\addlegendimage{color=\cl, only marks, mark = square*};%
				\noexpand\label{pl:cl:\i}
			}\colorlegend%
		}%
		\node[text width=0.6\textwidth, anchor=north west] at (axis cs:0.1125,0.8125) 
		{
			\foreach \i in {0,...,\LxsDim}%
			{%
				\pgfmathsetmacro{\Lx}{\Lxs[\i]}%
				\edef\colorlegend%
				{%
					\noexpand\ref{pl:cl:\i} $L_x=\noexpand\Lx\;$ %
				}\colorlegend
			}
		};
		\edef\markerlegend%
		{%
			\noexpand\addlegendimage{mark=\marker, color=black!60, thick};%
			\noexpand\label{pl:marker}
		}\markerlegend%
		\addlegendimage{dotted, color=black!60, thick};%
		\label{pl:marker:Ep}
		\node[text width=1.0\textwidth, align=left, anchor=north west] at (axis cs:0.1125,0.75)%
		{%
			\edef\markerlegend%
			{%
				\noexpand\ref{pl:marker} $t_z[J_y]=\tperpD\;$%
			}\markerlegend%
			\ref{pl:marker:Ep} $E_p\;$%
		};
		
		\foreach \i in {0,...,\LxsDim}%
		{%
			\pgfmathsetmacro{\cl}{\gpcolors[mod(\i,dim(\gpcolors))]}
			\pgfmathsetmacro{\Lx}{\Lxs[\i]}%
			\edef\plot%
			{%
				\noexpand\addplot%
				[%
					no marks,%
					color	=	\cl,%
					thick,%
					densely dotted,
				]%
					table%
					[%
						x expr = {(\noexpand\thisrowno{1}==0.7) ? \noexpand\thisrowno{0}/(2.0*\Lx) : NaN},%
						y expr = \noexpand\thisrowno{5},%
					]%
						{../data/results/allowed_critical_temperatures_Lx_\Lx_tperp_\tperpS.dat};%
			}\plot%
			\edef\plot%
			{%
					\noexpand\addplot%
					[%
						mark	=	\marker,%
						color	=	\cl,%
						thick,%
						restrict x to domain=0.13:1,%
					]%
						table%
						[%
							x expr = {(\noexpand\thisrowno{1}==0.7) ? \noexpand\thisrowno{0}/(2.0*\noexpand\Lx) : NaN},%
							y expr = {((\noexpand\thisrowno{3}<\noexpand\thisrowno{5})? ((\noexpand\thisrowno{4}<0.001)?\noexpand\thisrowno{3}: NaN) : NaN)},%
						]%
							{../data/results/allowed_critical_temperatures_Lx_\noexpand\Lx_tperp_\tperpS.dat};%
			}\plot %
			\edef\plot%
			{%
					\noexpand\addplot%
					[%
						mark	=	\marker,%
						color	=	\cl,%
						opacity	=	0.2,%
						only marks, %
						thick,%
						restrict x to domain=0.13:1,%
					]%
						table%
						[%
							x expr = {(\noexpand\thisrowno{1}==0.7) ? \noexpand\thisrowno{0}/(2.0*\noexpand\Lx) : NaN},%
							y expr = {((\noexpand\thisrowno{3}>\noexpand\thisrowno{5})? \noexpand\thisrowno{3} : NaN)},%
						]%
							{../data/results/allowed_critical_temperatures_Lx_\noexpand\Lx_tperp_\tperpS.dat};%
			}\plot %
		}%
		\coordinate (insetPos) at (axis description cs: 0.525,0.24);
	\end{axis}%
	\begin{axis}
	[	
		at = {(insetPos)},%
		width=0.5\textwidth,
		height=0.225\textheight,
		ymin = 0.075,
		ymax= 0.299,
		xlabel={$\nicefrac{t_z^2}{E_p}[J_y]$},
		ylabel={$T_c[J_y]$},
		xticklabels = {$0.06$, $0.08$, $0.1$, $0.12$},
		xtick = {0.06, 0.08, 0.1, 0.12}
	]
		\foreach \i in {0,...,\LxsDim}%
		{%
			\pgfmathsetmacro{\cl}{\gpcolors[mod(\i,dim(\gpcolors))]}
			\pgfmathsetmacro{\Lx}{\Lxs[\i]}%
			\edef\plot%
			{%
				\noexpand\addplot%
				[%
					mark	=	\marker,%
					color	=	\cl,%
					thick,%
				]%
					table%
					[%
						x expr = {(\noexpand\thisrowno{1}==0.7) ? (\tperpD*\tperpD)/\noexpand\thisrowno{5} : NaN},%
						y expr = {((\noexpand\thisrowno{3}<\noexpand\thisrowno{5})? ((\noexpand\thisrowno{4}<0.001)?\noexpand\thisrowno{3} : NaN ) : NaN)},%
					]%
						{../data/results/allowed_critical_temperatures_Lx_\noexpand\Lx_tperp_\tperpS.dat};%
			}\plot %
		}
	\end{axis}
\end{tikzpicture}%

%% file: figures/heuristic_spectrum.tex
\tikzsetnextfilename{heuristic_spectrum}%
\ifthenelse{\boolean{clickablePlot}}
{
	\tikzset{external/export next=false}%
	\def\txCriticalLxEight{73}%
	\def\txCriticalLxSixteen{104}%
	\def\txVisibleLxSixteenLeft{103}%
	\def\txVisibleLxSixteenRight{104}%
	\def\txCriticalLxThirtytwo{159}%
	\gdef\ocgs{} 
	\gdef\ocgsLeft{} 
	\gdef\ocgsRight{} 
	\foreach \Lx in {8, 16, 32}%
	{%
		\ifthenelse{\Lx = 8}%
		{%
			\def\L{Eight}%
			\def\txCritical{\txCriticalLxEight}%
		}%
		{%
			\ifthenelse{\Lx = 16}%
			{%
				\def\L{Sixteen}%
				\def\txCritical{\txCriticalLxSixteen}%
			}%
			{%
				\def\L{Thirtytwo}%
				\def\txCritical{\txCriticalLxThirtytwo}%
			}%
		}%
		\foreach \n in {30,31,...,170}%
		{%
			\ifthenelse{\n < \txCritical}%
			{%
				\xdef\ocgsLeft{\ocgsLeft\space ocg\L\n}%
				\xdef\ocgs{\ocgs\space ocg\L\n}%
			}%
			{%
				\xdef\ocgsRight{\ocgsRight\space ocg\L\n}%
				\xdef\ocgs{\ocgs\space ocg\L\n}%
			}%
		}%
	}%
	\def\allocgs{\ocgs}%
}{}%
\pgfdeclareplotmark{xocgA}%
{%
	\pgfsetstrokecolor{colorA}%
	\pgfpathmoveto{\pgfqpoint{-.70710678\pgfplotmarksize}{0\pgfplotmarksize}}%
	\pgfpathlineto{\pgfqpoint{.70710678\pgfplotmarksize}{0\pgfplotmarksize}}%
	\pgfpathmoveto{\pgfqpoint{0\pgfplotmarksize}{.70710678\pgfplotmarksize}}%
	\pgfpathlineto{\pgfqpoint{0\pgfplotmarksize}{-.70710678\pgfplotmarksize}}%
	\pgfmathtruncatemacro{\ocgnumber}{int(\pgfplotspointmetatransformed/5.025+1)}%
	\ifthenelse{\boolean{clickablePlot}}%
	{%
		\ifthenelse{\ocgnumber < \txCriticalLxEight}%
		{%
			\StrSubstitute{\ocgsLeft}{ocgEight\ocgnumber\space }{}[\ocgsLeftRemaining]%
			\node%
			[%
				actions ocg={ocgEight\ocgnumber}{}{\ocgsLeftRemaining},%
				draw opacity=0,%
				inner sep = 1.75pt,%
			]%
				{%
					\pgfusepathqstroke%
				};%
		}%
		{%
			\StrSubstitute{\ocgsRight}{ocgEight\ocgnumber\space }{}[\ocgsRightRemaining]%
			\node%
			[%
				actions ocg={ocgEight\ocgnumber}{}{\ocgsRightRemaining},%
				draw opacity=0,%
				inner sep = 1.75pt,%
			]%
				{%
					\pgfusepathqstroke%
				};%
		}%
	}%
	{\pgfusepathqstroke}%
}%
\pgfdeclareplotmark{xocgB}%
{%
	\pgfsetstrokecolor{colorB}%
	\pgfpathmoveto{\pgfqpoint{-.70710678\pgfplotmarksize}{0\pgfplotmarksize}}%
	\pgfpathlineto{\pgfqpoint{.70710678\pgfplotmarksize}{0\pgfplotmarksize}}%
	\pgfpathmoveto{\pgfqpoint{0\pgfplotmarksize}{.70710678\pgfplotmarksize}}%
	\pgfpathlineto{\pgfqpoint{0\pgfplotmarksize}{-.70710678\pgfplotmarksize}}%
	\pgfpathmoveto{\pgfqpoint{-.70710678\pgfplotmarksize}{-.70710678\pgfplotmarksize}}%
	\pgfpathlineto{\pgfqpoint{.70710678\pgfplotmarksize}{.70710678\pgfplotmarksize}}%
	\pgfpathmoveto{\pgfqpoint{-.70710678\pgfplotmarksize}{.70710678\pgfplotmarksize}}%
	\pgfpathlineto{\pgfqpoint{.70710678\pgfplotmarksize}{-.70710678\pgfplotmarksize}}%
	\ifthenelse{\boolean{clickablePlot}}%
	{%
		\pgfmathtruncatemacro{\ocgnumber}{int(\pgfplotspointmetatransformed/5.025+1)}%
		\ifthenelse{\ocgnumber < \txCriticalLxSixteen}%
		{%
			\StrSubstitute{\ocgsLeft}{ocgSixteen\ocgnumber\space }{}[\ocgsLeftRemaining]%
			\node%
			[%
				actions ocg={ocgSixteen\ocgnumber}{}{\ocgsLeftRemaining},%
				draw opacity=0,%
				inner sep = 1.75pt,%
			]%
				{%
					\pgfusepathqstroke%
				};%
		}%
		{%
			\StrSubstitute{\ocgsRight}{ocgSixteen\ocgnumber\space }{}[\ocgsRightRemaining]%
			\node%
			[%
				actions ocg={ocgSixteen\ocgnumber}{}{\ocgsRightRemaining},%
				draw opacity=0,%
				inner sep = 1.75pt,%
			]%
				{%
					\pgfusepathqstroke%
				};%
		}%
	}%
	{\pgfusepathqstroke}%
}%
\pgfdeclareplotmark{xocgC}%
{%
	\pgfsetstrokecolor{colorC}%
	\pgfpathmoveto{\pgfqpoint{-.70710678\pgfplotmarksize}{-.70710678\pgfplotmarksize}}%
	\pgfpathlineto{\pgfqpoint{.70710678\pgfplotmarksize}{.70710678\pgfplotmarksize}}%
	\pgfpathmoveto{\pgfqpoint{-.70710678\pgfplotmarksize}{.70710678\pgfplotmarksize}}%
	\pgfpathlineto{\pgfqpoint{.70710678\pgfplotmarksize}{-.70710678\pgfplotmarksize}}%
	\ifthenelse{\boolean{clickablePlot}}%
	{%
		\pgfmathtruncatemacro{\ocgnumber}{int(\pgfplotspointmetatransformed/5.025+1)}%
		\ifthenelse{\ocgnumber < \txCriticalLxThirtytwo}%
		{%
			\StrSubstitute{\ocgsLeft}{ocgThirtytwo\ocgnumber\space }{}[\ocgsLeftRemaining]%
			\node%
			[%
				actions ocg={ocgThirtytwo\ocgnumber}{}{\ocgsLeftRemaining},%
				draw opacity=0,%
				inner sep = 1.75pt,%
			]%
				{%
					\pgfusepathqstroke%
				};%
		}%
		{%
			\StrSubstitute{\ocgsRight}{ocgThirtytwo\ocgnumber\space }{}[\ocgsRightRemaining]%
			\node%
			[%
				actions ocg={ocgThirtytwo\ocgnumber}{}{\ocgsRightRemaining},%
				draw opacity=0,%
				inner sep = 1.75pt,%
			]%
				{%
					\pgfusepathqstroke%
				};%
		}%
	}%
	{\pgfusepathqstroke}%
}%
\begin{tikzpicture}%
	\begin{axis}%
	[%
		width			= 1.0\textwidth-16.8pt,%
		height			= 0.45\textheight,%
		xmin			= 0.3,%
		xmax			= 1.7,%
		ymin			= -0.3,%
		ymax			= 0.8,%
		xlabel			= {$t_x[J_y]$},%
		ylabel			= {$E_1+\frac{7}{8} t_x+\frac{L_x}{12}[J_y]$},%
		legend style	= {font=\scriptsize, at={(1,0.4775)}, anchor=east, draw = none, fill=none},%
		legend columns	= 1,%
	]%
		\addplot%
		[%
			only marks,%
			colorD,%
			mark			= *,%
			fill opacity	= 0.1,%
			draw opacity	= 0.2,%
		]%
			table%
			[%
				x expr	= {\thisrowno{0}},%
				y expr	= {\thisrowno{1}+(8-4*(8/32))*\thisrowno{0}+8/12},%
			]%
				{data/Ly_2/Lx_8/N_4/spectrum.dat};%
		\addlegendentry{$L_x=8$, ED}%
		\addplot%
		[%
			only marks,%
			scatter,%
			mark		= xocgA,%
			point meta	= {\thisrowno{0}},%
			restrict x to domain=0.3:1.7,
		]%
			table%
			[%
				x expr	= \thisrowno{0},%
				y expr	= {\thisrowno{2}+(8-4*(8/32))*\thisrowno{0}+8/12},%
			]%
				{data/Ly_2/Lx_8/N_4/energy_gss_fex_vs_tx.dat};%
		\addlegendentry{$L_x=8$}%
		\addplot%
		[%
			scatter,%
			only marks,%
			mark		= xocgB,%
			point meta	= {\thisrowno{0}},%
			restrict x to domain=0.3:1.7,
		]%
			table%
			[%
				x expr	= {\thisrowno{0}},%
				y expr	= {\thisrowno{2}+(16-4*(16/32))*\thisrowno{0}+16/12},%
			]%
				{data/Ly_2/Lx_16/N_8/energy_gss_fex_vs_tx.dat};%
		\addlegendentry{$L_x=16$}%
		\addplot%
		[%
			scatter,%
			only marks,%
			mark		=	xocgC,%
			point meta	=	{\thisrowno{0}},%
			restrict x to domain=0.3:1.7,
		]%
			table%
			[%
				x expr	= \thisrowno{0},%
				y expr	= {\thisrowno{2}+(32-4*(32/32))*\thisrowno{0}+32/12},%
			]%
				{data/Ly_2/Lx_32/N_16/energy_gss_fex_vs_tx.dat};%
		\addlegendentry{$L_x=32$}%
		\coordinate (inset_position_left) at (axis description cs:0.08,0.95);%
		\coordinate (inset_position_right) at (axis description cs:0.5,0.25);%
	\end{axis}%
	\ifthenelse{\boolean{clickablePlot}}%
	{%
		\foreach \Lx in {8, 16, 32}%
		{%
			\ifthenelse{\Lx = 8}%
			{%
				\def\L{Eight}%
				\def\N{4}%
				\def\txCritical{\txCriticalLxEight}%
			}%
			{%
				\ifthenelse{\Lx = 16}%
				{%
					\def\L{Sixteen}%
					\def\N{8}%
					\def\txCritical{\txCriticalLxSixteen}%
				}%
				{%
					\def\L{Thirtytwo}%
					\def\N{16}%
					\def\txCritical{\txCriticalLxThirtytwo}%
				}	%
			}%
			\foreach \txInt in {30,31,...,170}%
			{%
				\ifthenelse{\Lx = 16 \and \(\txInt = \txVisibleLxSixteenLeft \OR \txInt = \txVisibleLxSixteenRight\)}{\def\visibilityOfThisOcg{visible}}{\def\visibilityOfThisOcg{invisible}}%
				\edef\txP{\fpeval{round(\txInt/100,2)}}%
				\StrSubstitute{\txP}{.}{p}[\txS]%
				\StrSubstitute{\txS}{p}{.}[\txP]%
				\ifthenelse{\txInt < \txCritical}%
				{%
					\def\InsetPosition{inset_position_left}%
				}%
				{%
					\def\InsetPosition{inset_position_right}%
				}%
				\begin{axis}%
				[%
					name 		= ocgplot,%
					ocg 		= {name=ocg\L\txInt, ref=ocg\L\txInt, status=\visibilityOfThisOcg},%
					inner sep	= 1pt,%
					at			= {(\InsetPosition)},%
					anchor		= north west,%
					width		= 0.49\textwidth,%
					height		= 0.14\textheight,%
					xlabel		= {\tiny Site $j$},%
					ylabel		= {\tiny $\braket{\hat N_j}$},%
					title		= {\tiny $L_x=\Lx$, $t_x=\txP$},%
					axis background/.style			= {fill=white, fill opacity=0.9},%
					title style						= {yshift=-0.35em, xshift=-1em, fill=white, name=title},%
					every tick label/.append style	= {font=\tiny, fill=white, fill opacity=0.9},%
					xlabel style					= {name=xlabel, fill=white, fill opacity=0.9},%
					legend style 					= {draw=none, fill=none, anchor=south, at = {(0.5,0)}, inner sep = 1pt,},%
				]%
					\addplot%
					[%
						color				= blue,%
						mark				= x,%
						unbounded coords	= jump,%
					]%
						table%
						[%
							x expr 	= \thisrowno{0},%
							y expr 	= \thisrowno{1},%
						]%
							{data/Ly_2/Lx_\Lx/N_\N/S_0/t_x_\txS/Jz_x_0p047/D_x_0p047/V_x_0p047/t_y_0p0/Jz_y_1p0/D_y_1p0/V_y_1p0/density_chimax_1024.dat};%
					\addlegendentry{\tiny Ground state};%
					\addplot%
					[%
						color				= black,%
						mark				= x,%
						unbounded coords	= jump,%
					]%
						table%
						[%
							x expr 	= \thisrowno{0},%
							y expr 	= \thisrowno{1},%
						]%
							{data/Ly_2/Lx_\Lx/N_\N/S_0/t_x_\txS/Jz_x_0p047/D_x_0p047/V_x_0p047/t_y_0p0/Jz_y_1p0/D_y_1p0/V_y_1p0/density_chimax_1024.datfirst_excited_state};%
					\addlegendentry{\tiny First excited state};%
				\end{axis}%
			}%
		}%
		\node%
		[%
			hide ocg	= {\ocgs},%
			fit 		= (ocgplot),%
			yshift		= -2.5em,%
			inner sep	= 1.5em,%
			xshift		= -1em%
		]%
			{};%
	}%
	{%
		\def\Lx{16}%
		\def\N{8}%
		\def\txP{1.03}%
		\def\txS{1p03}%
		\begin{axis}%
		[%
			inner sep	= 1pt,%
			at			= {(inset_position_left)},%
			anchor		= north west,%
			width		= 0.49\textwidth,%
			height		= 0.14\textheight,%
			xlabel		= {\tiny Site $j$},%
			ylabel		= {\tiny $\braket{\hat N_j}$},%
			title		= {\tiny $L_x=\Lx$, $t_x=\txP$},%
			axis background/.style			= {fill=white, fill opacity=0.9},%
			title style						= {yshift=-0.35em, xshift=-1em, fill=white, name=title, fill opacity=0.8},%
			every tick label/.append style	= {font=\tiny, fill=white, fill opacity=0.8},%
			xlabel style					= {name=xlabel, fill=white, fill opacity=0.8},%
			legend style 					= {draw=none, fill=none, anchor=south, at = {(0.5,0)}, inner sep = 1pt,},%
			axis background/.style			= {fill=white, fill opacity=0.65,},%
		]%
			\addplot%
			[%
				color				= blue,%
				mark				= x,%
				unbounded coords	= jump,%
			]%
				table%
				[%
					x expr 	= \thisrowno{0},%
					y expr 	= \thisrowno{1},%
				]%
					{data/Ly_2/Lx_\Lx/N_\N/S_0/t_x_\txS/Jz_x_0p047/D_x_0p047/V_x_0p047/t_y_0p0/Jz_y_1p0/D_y_1p0/V_y_1p0/density_chimax_1024.dat};%
			\addlegendentry{\tiny Ground state};%
			\addplot%
			[%
				color				= black,%
				mark				= x,%
				unbounded coords	= jump,%
			]%
				table%
				[%
					x expr 	= \thisrowno{0},%
					y expr 	= \thisrowno{1},%
				]%
				{data/Ly_2/Lx_\Lx/N_\N/S_0/t_x_\txS/Jz_x_0p047/D_x_0p047/V_x_0p047/t_y_0p0/Jz_y_1p0/D_y_1p0/V_y_1p0/density_chimax_1024.datfirst_excited_state};%
			\addlegendentry{\tiny First excited state};%
		\end{axis}%
		\def\Lx{16}%
		\def\N{8}%
		\def\txP{1.04}%
		\def\txS{1p04}%
		\begin{axis}%
		[%
			inner sep	= 1pt,%
			at			= {(inset_position_right)},%
			anchor		= north west,%
			width		= 0.49\textwidth,%
			height		= 0.14\textheight,%
			xlabel		= {\tiny Site $j$},%
			ylabel		= {\tiny $\braket{\hat N_j}$},%
			title		= {\tiny $L_x=\Lx$, $t_x=\txP$},%
			axis background/.style			= {fill=white, fill opacity=0.9},%
			title style						= {yshift=-0.35em, xshift=-1em,fill=white, name=title, fill opacity=0.8},%
			every tick label/.append style	= {font=\tiny, fill=white, fill opacity=0.8},%
			xlabel style					= {name=xlabel,fill=white, fill opacity=0.8},%
			legend style 					= {draw=none, fill=none, anchor=south, at = {(0.5,0)}, inner sep = 1pt,},%
			axis background/.style			= {fill=white, fill opacity=0.65,},%
		]%
			\addplot%
			[%
				color				= blue,%
				mark				= x,%
				unbounded coords	= jump,%
			]%
				table%
				[%
					x expr 	= \thisrowno{0},%
					y expr 	= \thisrowno{1},%
				]%
				{data/Ly_2/Lx_\Lx/N_\N/S_0/t_x_\txS/Jz_x_0p047/D_x_0p047/V_x_0p047/t_y_0p0/Jz_y_1p0/D_y_1p0/V_y_1p0/density_chimax_1024.dat};%
			\addlegendentry{\tiny Ground state};%
			\addplot%
			[%
				color				= black,%
				mark				= x,%
				unbounded coords	= jump,%
			]%
				table%
				[%
					x expr 	= \thisrowno{0},%
					y expr 	= \thisrowno{1},%
				]%
				{data/Ly_2/Lx_\Lx/N_\N/S_0/t_x_\txS/Jz_x_0p047/D_x_0p047/V_x_0p047/t_y_0p0/Jz_y_1p0/D_y_1p0/V_y_1p0/density_chimax_1024.datfirst_excited_state};%
			\addlegendentry{\tiny First excited state};%
		\end{axis}%
	}%
\end{tikzpicture}%

%% file: figures/R_vs_K_c.tex
\tikzsetnextfilename{R_vs_K_c}%
\begin{tikzpicture}%
	\begin{axis}%
	[%
			width = 1.0\textwidth-8.32pt,%
			height = 0.3\textheight,%
			xlabel = {$K_c$},%
			ylabel = {$R(K_c)$},%
	]%
		\addplot%
		[%
			thick,%
		]%
			table%
			[%
				x expr = {\thisrowno{0}},%
				y expr = {\thisrowno{1}}
			]%
				{../data/ratio_vs_k_c.dat};%
	\end{axis}
\end{tikzpicture}%